\documentclass[fleqn,usenatbib]{mnras}

\usepackage{mathptmx}
\usepackage{comment}

\usepackage[T1]{fontenc}
\usepackage{ae,aecompl}

\usepackage{longtable}
\usepackage{placeins}
\usepackage{multicol}
\usepackage{subfigure}

\usepackage{graphicx}	
\usepackage{amsmath}	
\usepackage{amssymb}	
\usepackage{siunitx} 

\title[Outflows in NGC\,1275]{Ionized and hot molecular outflows in the inner 500 pc of NGC\,1275}

\author[R. A. Riffel et al.]{Rogemar A. Riffel,$^{1,2}$\thanks{E-mail: rogemar@ufsm.br (RAR)}
Thaisa Storchi-Bergmann,$^{3}$ Nadia L. Zakamska,$^{2}$ \and Rog\'erio Riffel$^{3}$
\\
$^{1}$Departamento de F\'isica, CCNE, Universidade Federal de Santa Maria, 97105-900, Santa Maria, RS, Brazil\\
$^{2}$Department of Physics \& Astronomy, Johns Hopkins University, Bloomberg Center, 3400 N. Charles St, Baltimore, MD 21218, USA\\
$^{3}$Universidade Federal do Rio Grande do Sul, IF, CP 15051, Porto Alegre 91501-970, RS, Brazil
}

\date{Accepted XXX. Received YYY; in original form ZZZ}

\pubyear{2020}

\begin{document}
\label{firstpage}
\pagerange{\pageref{firstpage}--\pageref{lastpage}}
\maketitle

\begin{abstract}
The role of feedback from Active Galactic Nuclei (AGN) in the evolution of galaxies is still not not fully understood, mostly due to the lack of observational constraints in the multi-phase gas kinematics on the ten to hundred parsec scales. We have used the Gemini Near-infrared Integral Field Spectrograph (NIFS) to map the molecular and ionized gas kinematics in the inner 900$\times$900 pc$^2$ of the Seyfert galaxy NGC\,1275 at a spatial resolution of $\sim$70 pc. From the fitting of the CO absorption bandheads in the K-band, we derive a stellar velocity dispersion of $265\pm26$ km\,s$^{-1}$, which implies a black hole mass of  $M_{\rm SMBH}=1.1^{+0.9}_{-0.5}\times10^9$\,M$_\odot$. We find hot ($T\gtrsim1000$ K) molecular  and ionized outflows with velocities of up to 2\,000 km s$^{-1}$ and mass outflow rates of $2.7\times10^{-2}\,{\rm M_\odot}$\,yr$^{-1}$ and $1.6\,{\rm M_\odot}$\,yr$^{-1}$, respectively, in each of these gas phases. The kinetic power of the ionized outflows corresponds to only 0.05 per cent of the luminosity of the AGN of NGC\,1275, indicating that they are not powerful enough to provide significant AGN feedback, but may be effective in redistributing the gas in the central region of the galaxy. The AGN driven outflows seem to be responsible for the shocks necessary to produce the observed H$_2$ and [Fe\,{\sc ii}] line emission.

\end{abstract}




\section{Introduction}

  Perseus is the X-ray brightest known galaxy cluster \citep{forman72} and a cool core cluster prototype \citep{crawford99}. It shows a rich emission structure, presenting multiple X-ray cavities, ripples, shocks and filaments  \citep[e.g.][]{fabian00,fabian03,fabian06,sanders16}. The brightest member of the cluster is the giant elliptical galaxy NGC\,1275   \citep[Perseus A, 3C84, ][]{sandage81}, located at its center, at a distance of 62.5 \,Mpc \citep{tully13}. NGC\,1275 hosts a Seyfert 1.5 nucleus \citep{veron06} and has strong infrared emission with a luminosity of  $\log(L_{\rm IR}/L_{\odot}) = 11.20$, consistent with a Luminous Infra-Red Galaxy  \citep{sanders03}. 
   It presents collimated  radio emission  along the Position Angle PA=160/340$^\circ$ extending to up to 30$^{\prime\prime}$ from its nucleus \citep{pedlar90}, a sub-pc scale intermittent radio jet \citep[e.g.][]{nagai10,suzuki12} and variable high-energy  $\gamma$-ray emission \citep{abdo09}. 
 
 The most intriguing feature of NGC\,1275 is the kpc scale filamentary structure seen both in molecular \citep{salome06,hatch05,lim08,ho09b} and ionized gas \citep{kent79,conselice01,fabian08}. Recently, \citet{gendron18} presented integral-field spectra obtained with the instrument SITELLE at the Canada France Hawaii Telescope, that provide a detailed mapping of the ionization and kinematics of the large scale filamentary nebulae surrounding NGC\,1275, covering an area of 80$\times$40\,kpc$^2$ (3.8$\times$2.6\,arcmin$^2$) with a spatial resolution of $\sim$0\farcs3. These observations reveal that the kinematics of the filaments do not show any signature of ordered motions, interpreted by the authors as an evidence that the filaments are not consistent with uniform inflows or outflows. Their analysis exclude the inner 6$^{\prime\prime}$, where the line emission originates from the gas ionized by the central Active Galactic Nucleus (AGN).  
 
 Besides strong optical emission lines from the ionized gas, the nucleus of NGC\,1275 also presents strong molecular hydrogen emission, seen in the mid \citep[e.g.][]{armus07,lambrides19} and near-infrared \citep[e.g.][]{krabbe00,wilman05,rogerio06} spectra.  NGC\,1275 presents H$_2$ emission excess in the mid infrared in comparison with other nearby galaxies, with $\log$[H$_2$\,S(3)$\lambda$9.665\,$\mu$m/PAH$\lambda$11.3\,$\mu$m]$\approx-0.18$ \citep{lambrides19} and  high [O\,{\sc i}]$\lambda6300$\,\AA\ velocity dispersion ($\sim$360\,km\,s$^{-1}$) and [O\,{\sc i}]$\lambda6300$\,\AA/H$\alpha$ ($\sim$1.75) \citep{gavazzi13}. 
 The simultaneous enhancement in H$_2$ emission and in [O\,{\sc i}] velocity dispersion make this object a likely host of molecular outflows \citep{rogemar19} and therefore a unique nearby laboratory to probe AGN feedback. The near-IR spectral region is less affected by dust extinction than the optical bands and presents emission lines from distinct gas phases, allowing the mapping of the multi-phase gas kinematics and distribution. Here, we present new Gemini Near-Infrared Integral Field Spectrograph (NIFS) J- and K-band data used to used to unravel the gas kinematics of the central $900\times900\,{\rm pc^2}$ of NGC\,1275.     
 
 Gemini NIFS observations in the H and K band have already been used to study the gas excitation and kinematics of central region of NGC\,1275 
 \citep{nifs13}. These data reveal a circum-nuclear disk in the inner 100 pc of the galaxy oriented along the position angle PA=68$^\circ$, which is modeled by a Keplerian rotation arround a central mass of 8$^{+7}_{-2}\times10^8 {\rm M_\odot}$, taken as the mass of the supermassive black hole. Recent Atacama Large Millimeter Array (ALMA) CO(2--10) observations show the cold counterpart of the disk, as well as fast outflows (300--600 km\,s$^{-1}$) detected as absorption features in the  HCN (3--2) and HCO$^+$ (3--2) spectra \citep{nagai19}. 
  Based on single Gaussian fit of the H$_2$ and [Fe\,{\sc ii}] emission-line profiles, \citet{nifs13} find a redshifted H$_2$ structure to the south-west of the nucleus, possibly falling into the center. Another redshifted blob is seen in both H$_2$ and [Fe\,{\sc ii}] at 1\farcs2 north-west of the nucleus, coincident with the orientation of the radio jet and tentatively interpreted as an evidence of jet-cloud interaction. Based on line ratios obtained from integrated spectra, \citet{nifs13} also conclude that most likely the H$_2$ emission in the inner region of NGC\,1275 is produced by shocks and the [Fe\,{\sc ii}] emission is consistent with being originated by X-ray heating, but they could not rule out a possible contribution by shocks. 
  
  As noticed by \citet{nifs13}, the near-IR emission-line profiles of the central region are complex. In this work, we analyse the gas kinematics and distribution based on non-parametric measurements and multi-Gaussian components fits, which allows us to better  disentangle the distinct kinematic components. In addition, the J-band spectra allow us to better map the AGN ionization structure than the previous available data, fundamental to understand the origin of the gas emission and the role of the AGN. 
  
  This paper is organized as follows. Section 2 presents a description of the data, its reduction and analysis procedures. In Sec. 3, we present two-dimensional maps of emission-line fluxes, line ratio and kinematics, which are discussed in Sec. 4. Finally, Sec. 5 presents our conclusions.  
 
\section{Observations and data reduction}\label{sec:data}

The Gemini NIFS \citep{mcgregor03} is an integral field spectrograph optimized to operate with the ALTtitude conjugate Adaptive optics for the InfraRed (ALTAIR). It has a square field of view of $\approx3\farcs0\times3\farcs0$, divided into 29 slices with an angular sampling of 0$\farcs$103$\times$0$\farcs$042. The observations of NGC\,1275 were done in October 22, 2019 in the queue mode under the project GN-2019A-Q-106. For the J-band, we use the J$_-$G5603 grating and the ZJ$_-$G0601 filter with a resolving power of $R\approx6040$, while for the K-band we use the K$_-$G5605 grating and HK$_-$G0603 filter, resulting in $R\approx5290$.  

The observations followed the standard Object-Sky-Object dither sequence, with off-source sky positions at 1$^\prime$ from the galaxy, and individual exposure times of
470\,s. The J-band spectra are centred at 1.25 $\mu$m, covering the spectral range from 1.14 to 1.36\,$\mu$m. The K-band data are centred at 2.20\,$\mu$m and cover the 2.00--2.40\,$\mu$m spectral region. The total exposure time at each band was 47 min.

The data reduction was accomplished using tasks contained in the {\sc nifs}
package which is part of {\sc gemini iraf} package, as well as generic {\sc
iraf} tasks. The data reduction followed the standard procedure, which includes  
the  trimming of the images, flat-fielding, cosmic ray rejection, sky subtraction, wavelength and s-distortion calibrations.
In order to remove telluric absorptions from the galaxy spectra we observed the telluric standard star HIP\,15925
 just after the J-band observations of the galaxy and  HIP\,10559 just before the K-band observations.
The galaxy spectra was divided by the normalized spectrum of the tellurc standard star using the {\sc nftelluric} task of 
the {\sc nifs.gemini.iraf} package. The galaxy spectra were flux calibrated by interpolating a blackbody function 
to the spectrum of the telluric standard and the J and  K-bands data cubes were constructed with an 
angular sampling of 0\farcs05$\times$0\farcs05 for each individual exposure. The individual data cubes 
were combined using a sigma clipping algorithm in order to eliminate bad pixels and remaining cosmic rays by mosaicing the dithered spatial positions. Then, we apply the method described in \citet{davies07} to remove residual OH airglow emission from near-infrared spectra and finally, we use a Butterworth spatial band-pass filter to remove high-frequency noise, by adopting a cutoff frequency of 0.25 Ny.

The angular resolution of the K-band data cube is 0\farcs22$\pm$0\farcs03 (67$\pm$9\,pc at the galaxy), as measured from the full width at half maximum (FWHM) of the flux distribution of the Br$\gamma$ broad component emission. For the J-band emission, the final data cube has an angular resolution of 0\farcs25$\pm$0\farcs02 (76$\pm$6\,pc at the galaxy), estimated from the flux distribution of the broad component of Pa$\beta$. The resulting velocity resolution is 43$\pm$5\,km\,s$^{-1}$ in the J-band and 47$\pm$5\,km\,s$^{-1}$ in the K-band, measured from the FWHM of typical emission lines of the Ar and ArXe lamps spectra used to wavelength calibrate the J and K band spectra, respectively.

\begin{figure*}
    \centering
    \includegraphics[width=0.75\textwidth]{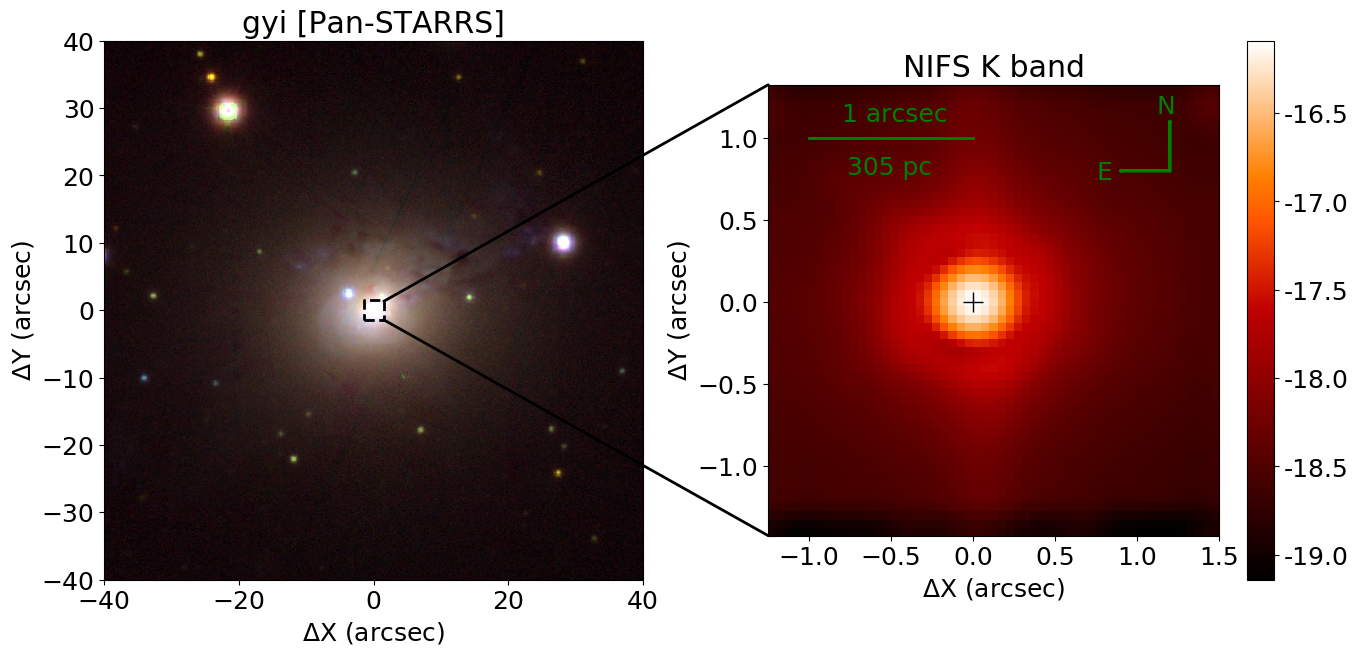}
    \includegraphics[width=0.75\textwidth]{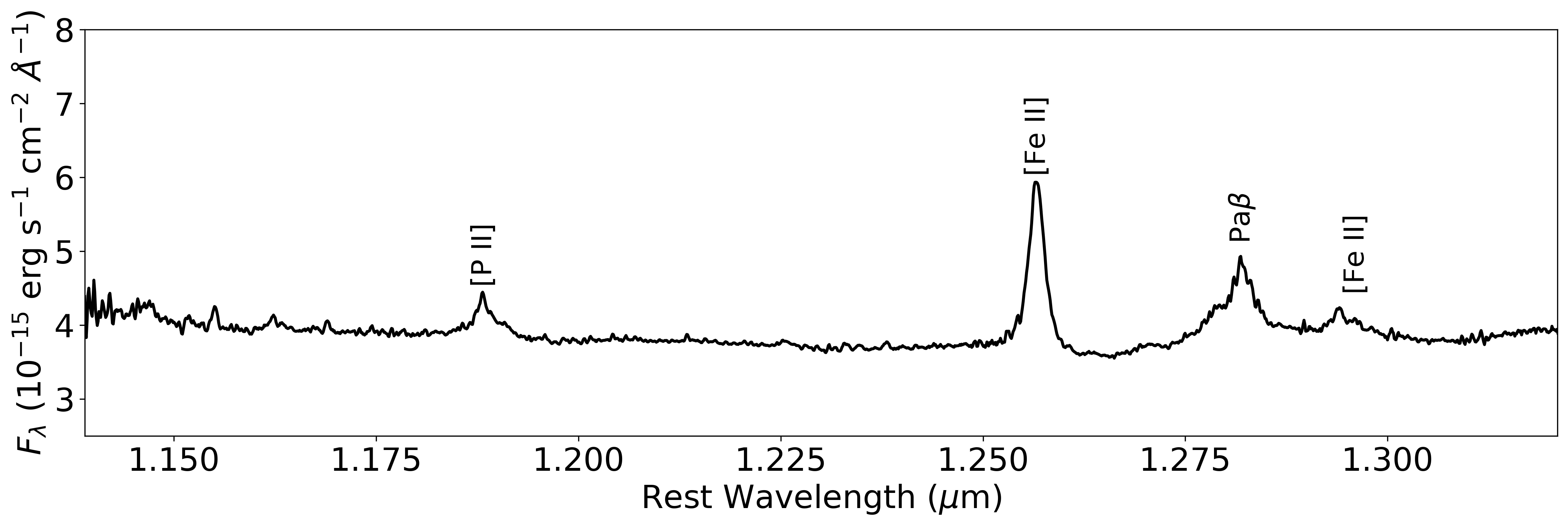}
    \includegraphics[width=0.75\textwidth]{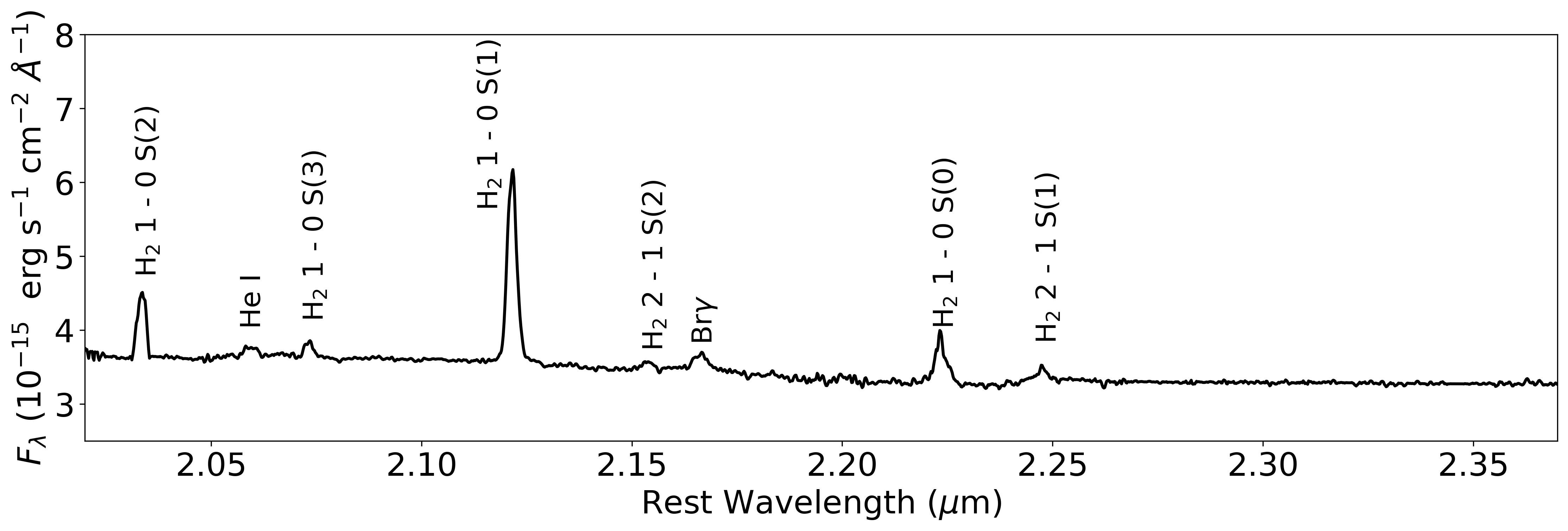}
    \caption{Top left: $gyi$ image of NGC\,1275 from Pan-STARRS data archive \citep{chambers16,flewelling16}. Top right: NIFS K-band continuum. The color bar show the fluxes in logarithmic units of erg\,s$^{-1}$\,cm$^{-1}$\,\AA$^{-1}$\,spaxel$^{-1}$. The bottom panels show the J and K band spectra of NGC\,1275 within an aperture of 1\farcs0 radius, centred at the position of the peak of the continuum.  }
    \label{spectra}
\end{figure*}

The top-left panel of Figure\,\ref{spectra} presents a large scale $gyi$ composed image of NGC\,1275 obtained from the Pan-STARRS data archive \citep{chambers16,flewelling16} and the K-band continuum image, obtained by computing the average flux of the NIFS data cube at each spaxel, is presented in the top-right panel. The bottom panels show the J and K-band spectra integrated within an aperture of 1\farcs0 radius centred at the location of the peak of the continuum emission. The main emission lines are identified: [P\,{\sc ii}]$\lambda$1.1886\,$\mu$m, [Fe\,{\sc ii}]$\lambda$1.1886\,$\mu$m 1.2570, Pa$\beta\,\lambda$1.2822\,$\mu$m, [Fe\,{\sc ii}]$\lambda$1.2964\,$\mu$m, H$_2$\,$\lambda$2.0338\,$\mu$m, He\,{\sc i}$\lambda$2.0587\,$\mu$m, H$_2$\,$\lambda$2.0735\,$\mu$m,  H$_2$\,$\lambda$2.1218\,$\mu$m, H$_2$\,$\lambda$2.1542\,$\mu$m, Br$\gamma$\,$\lambda$2.16612\,$\mu$m, H$_2$\,$\lambda$2.2233\,$\mu$m, H$_2$\,$\lambda$2.2477\,$\mu$m.

\section{Measurements}\label{sec:measurements}

\subsection{Non-parametric properties}

We construct two-dimensional maps for the emission-line flux distributions, flux ratios and for the width in velocity space which encompasses 80 per cent of the flux of the emission lines ($W_{\rm 80}$). For asymmetric, non-Gaussian emission-line profiles, the $W_{\rm 80}$ is a better estimate of the width of the line profiles in galaxies than the FWHM and it is widely used to identify outflows in nearby and distant galaxies \citep{zakamska14,wylezalek17,wylezalek20}. We measure the emission-line fluxes and W$_{\rm 80}$ using the {\it IFSCube} python package\footnote{https://ifscube.readthedocs.io}. The Pa$\beta$ profile clearly presents a very broad component ($\sigma\sim2600$ km\,s$^{-1}$, from the Broad Line Region -- BLR) and thus, first we fit the line profile by two-Gaussian components and subtract the flux of the broad component. A weak  emission from the BLR is also seen in Br$\gamma$ and we follow the same procedure. The fluxes are integrated in the spectral window that encompasses 5$\sigma$ to each side of the peak of the line, as measured from the narrow line component.

\begin{figure*}
    \centering
\includegraphics[width=0.23\textwidth]{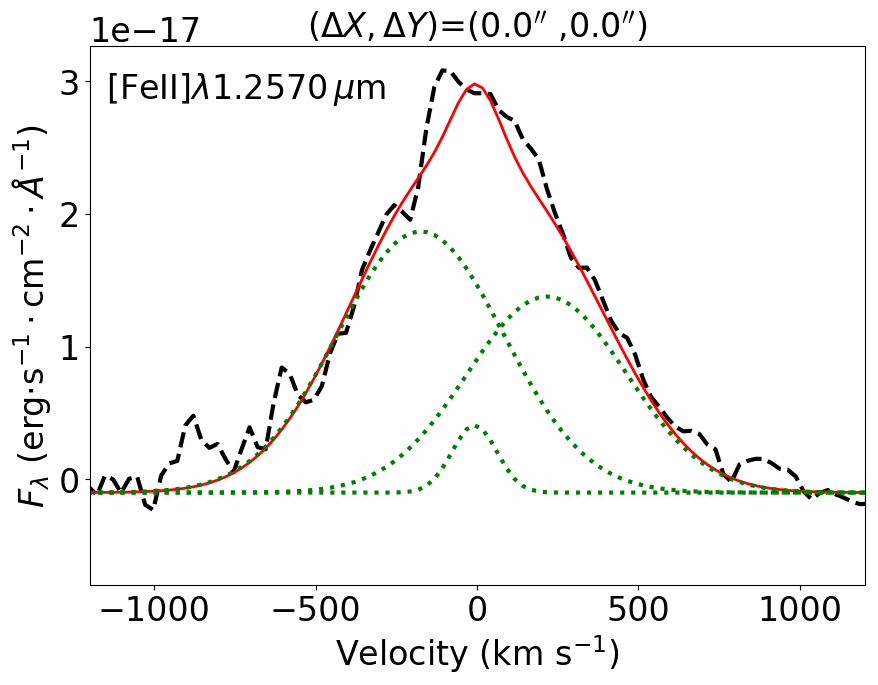}
\includegraphics[width=0.23\textwidth]{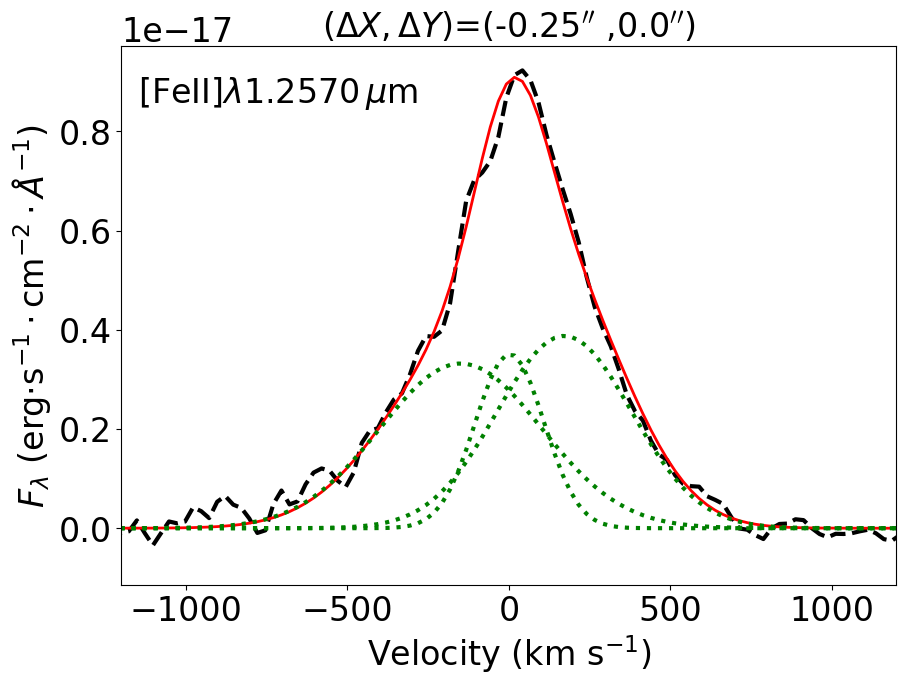}
\includegraphics[width=0.23\textwidth]{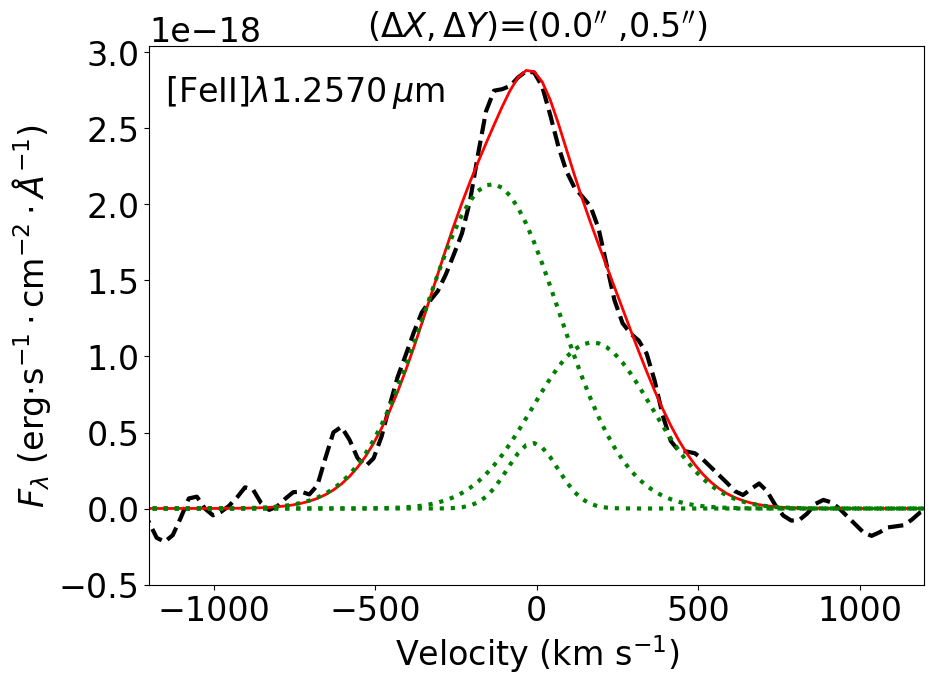}
\includegraphics[width=0.23\textwidth]{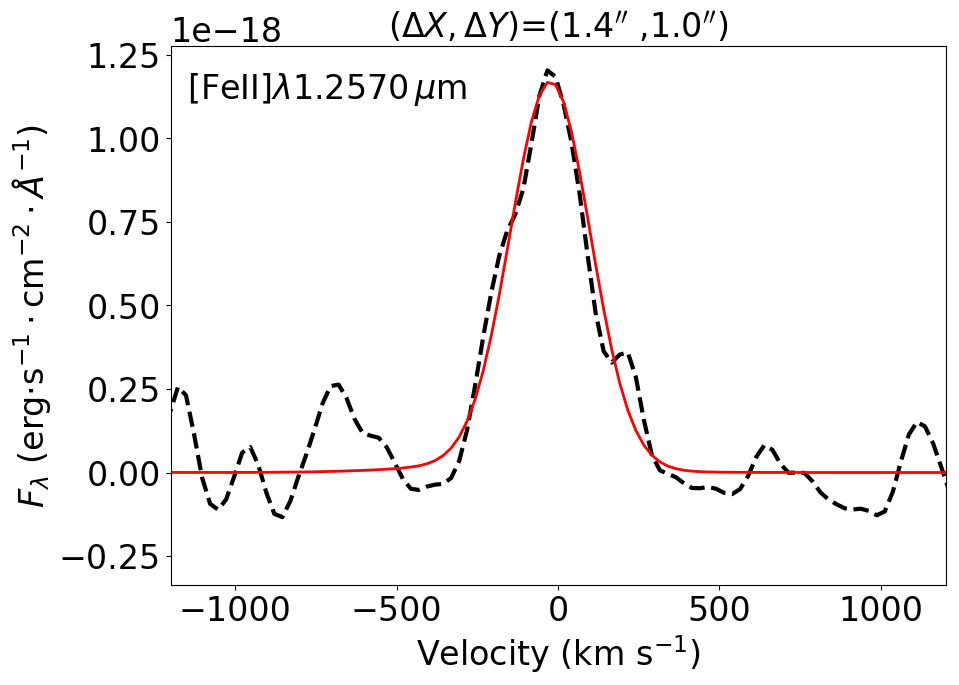}
\includegraphics[width=0.23\textwidth]{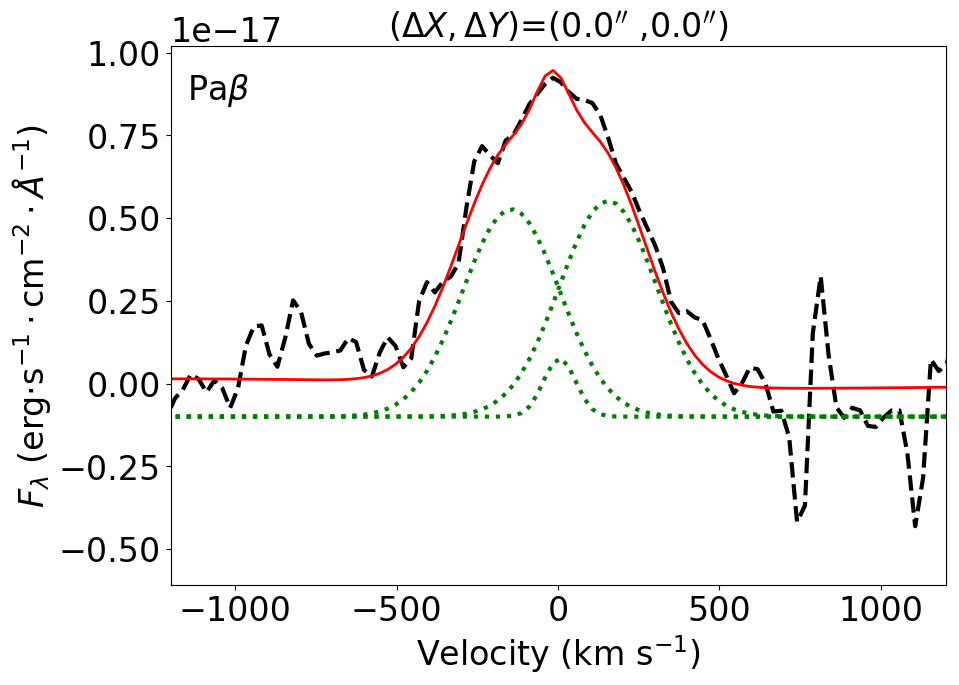}
\includegraphics[width=0.23\textwidth]{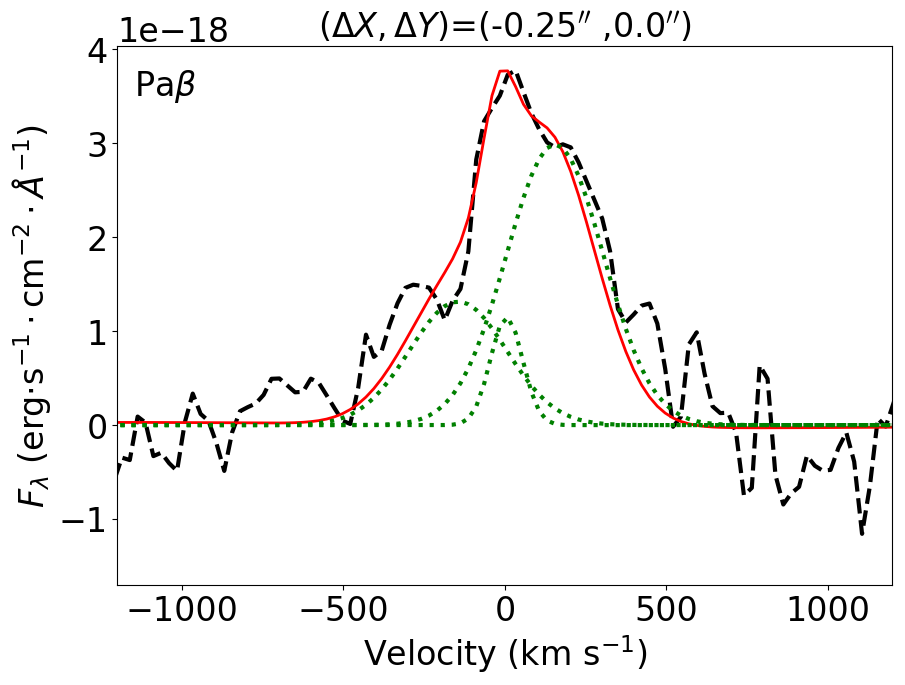}
\includegraphics[width=0.23\textwidth]{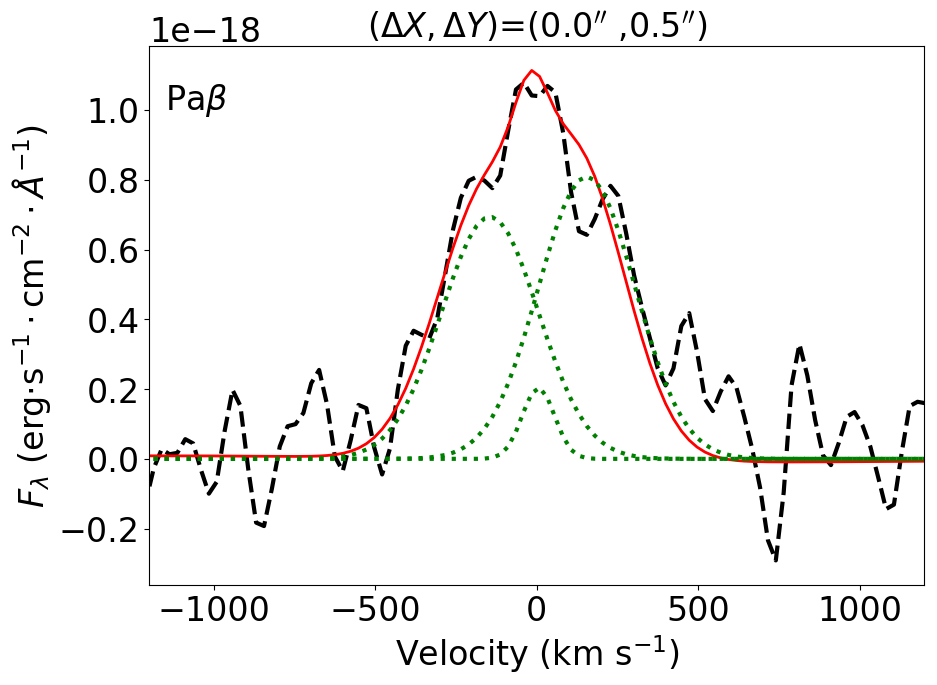}
\includegraphics[width=0.23\textwidth]{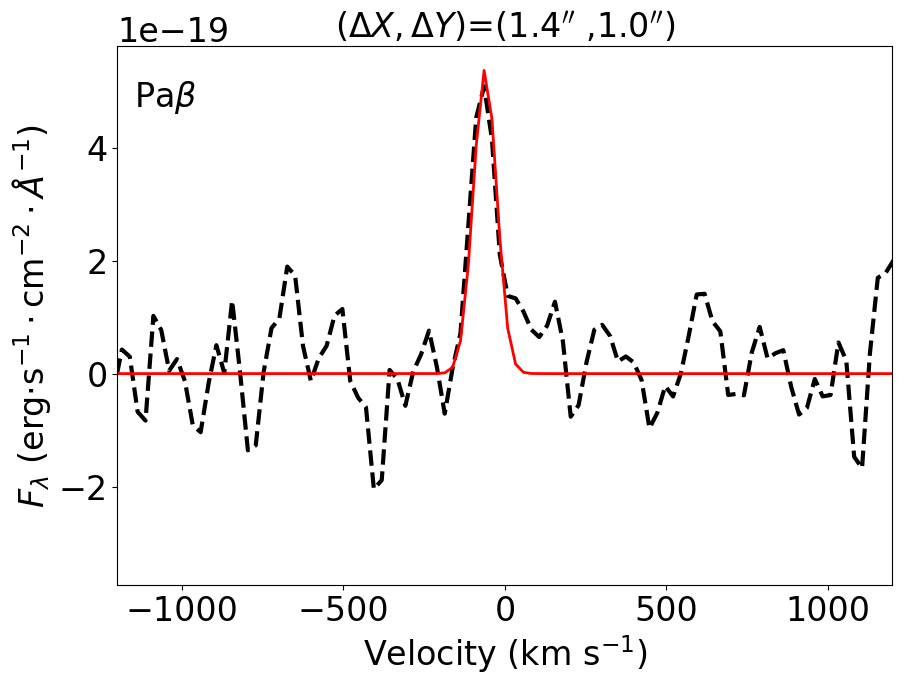}
\includegraphics[width=0.23\textwidth]{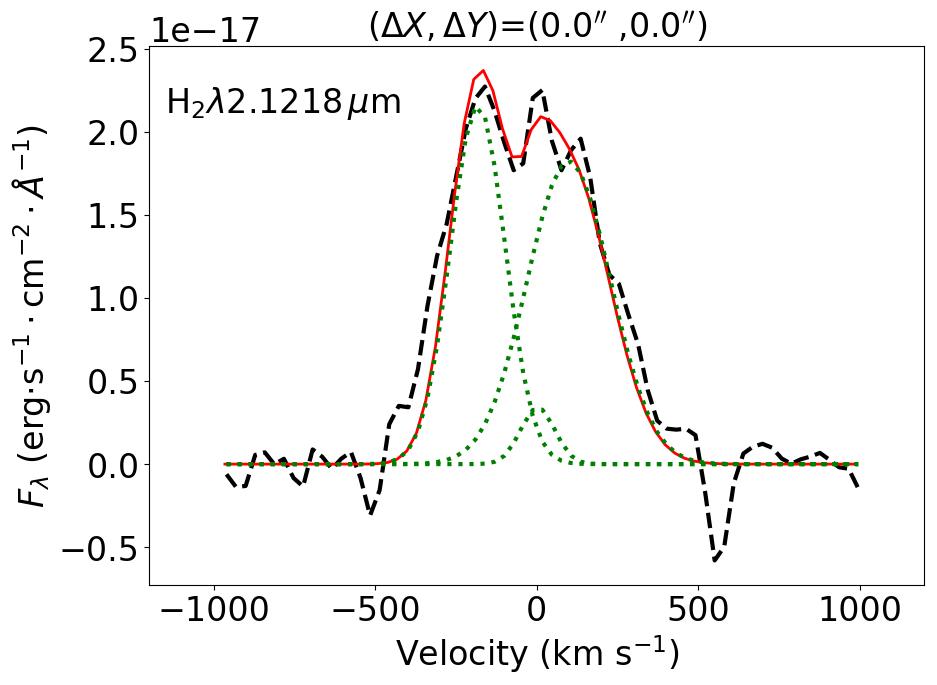}
\includegraphics[width=0.23\textwidth]{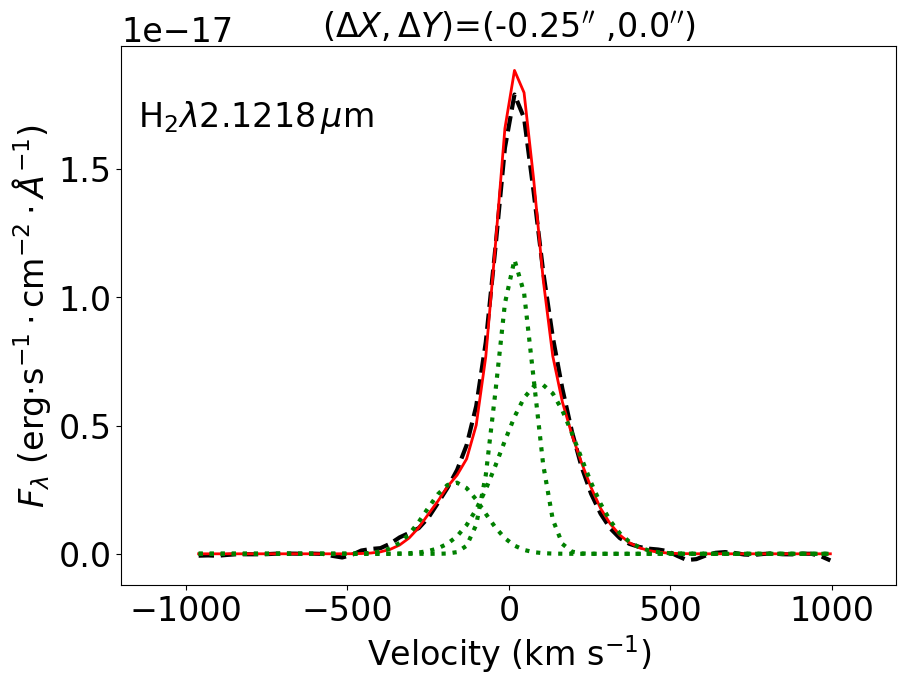}
\includegraphics[width=0.23\textwidth]{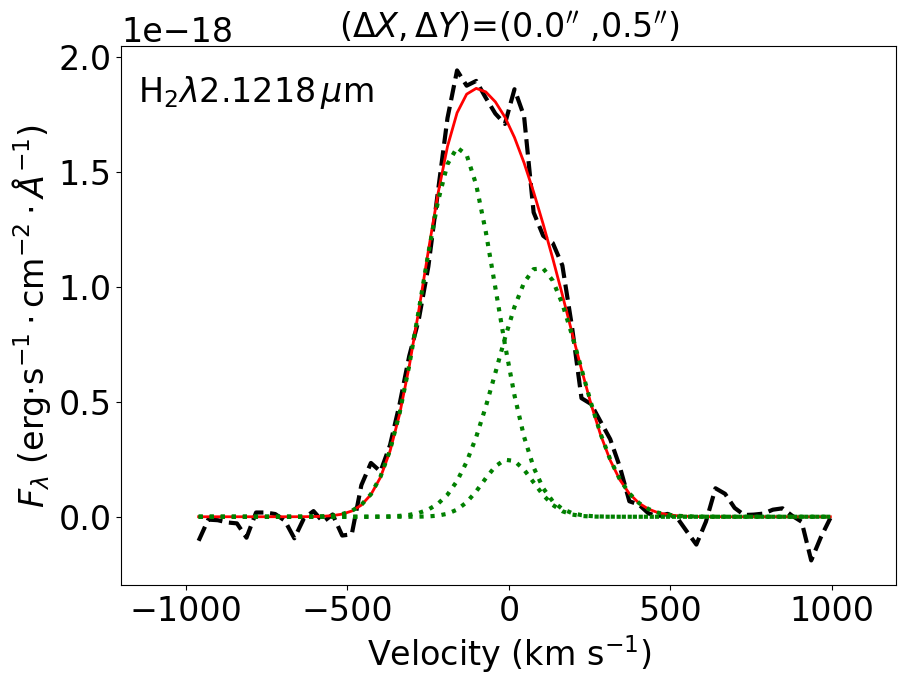}
\includegraphics[width=0.23\textwidth]{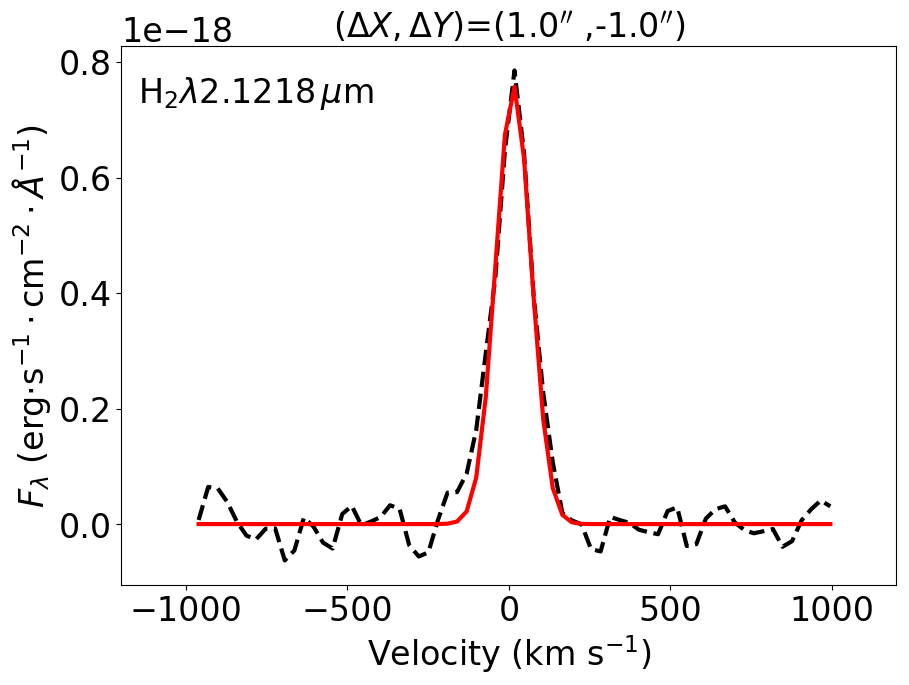}
    \caption{Examples of fitting of the [Fe{\sc ii}]$\lambda1.2570\,\mu$m (first row), Pa$\beta$ (second row) and H$_2\lambda2.1218\,\mu$m (third row) emission line profiles at the position ($\Delta X,\Delta Y$) relative to the nucleus identified in the top of each panel. The  continuum-subtracted observed profiles  are shown as dashed black lines, the fits are in red and the individual Gaussian components are shown as green dotted lines. For the Pa$\beta$, besides the continuum emission, the broad component (from the broad line region) has also been subtracted in the two left-most plots.
}
    \label{linefits} 
\end{figure*}

\subsection{Emission-line fitting}\label{ELfits}

By fitting a single Gaussian to each emission-line profile, we recover the maps presented by \citet{nifs13} for the K-band emission lines and we find that the [Fe\,{\sc ii}]$\lambda$1.2570\,$\mu$m shows similar maps to those of the H-band presented by these authors. We do not show  maps based on single Gaussian fits, as they do not add to the discussion already presented in \citet{nifs13}. However, the emission-line profiles in the inner 3$^{\prime\prime}\times{\rm 3}^{\prime\prime}$ (900$\times$900\,pc$^2$) of NGC\,1275 are complex and in most locations they are not well modeled neither by a single Gaussian, nor by Gauss-Hermite series \citep{marel93,profit}. Thus, we fit the emission-line profiles by multiple-Gaussian components using the {\it IFSCube} python package. 

We allow the fit of up to three Gaussian components to each emission-line profile at all positions. The J and K-band spectra are fitted separately and all parameters are kept free, as the line profiles of distinct species clearly show distinct kinematic components (Fig.~\ref{linefits}).  We set the {\it refit} parameter of the {\sc ifscube} code to use the best-fit parameters from spaxels located at distances smaller than 0\farcs15 as the initial guess for the next fit. Thus, the fitting routine automatically chooses the best number of components and the initial guesses for the fit of a determined line profile at a specific spaxel. For example, if the amplitude of one of the Gaussians is zero in the neighboring spaxels, the code will return the parameters of the two remaining Gaussian curves and so on. As initial guesses for the first fit of each emission line we use the parameters of a three Gaussian fit for the nuclear spaxel using the {\sc iraf} {\sc splot} task.
For Pa$\beta$ and Br$\gamma$, we include an additional very broad component ($\sigma\sim1000$ km\,s$^{-1}$) to take into account the emission from the BLR.

In Figure~\ref{linefits} we show examples of the fits of the [Fe\,{\sc ii}]$\lambda1.2570\,\mu$m (top row), Pa$\beta$ (middle row) and H$_2\lambda2.1218\,\mu$m (bottom row) for four different positions. As can be seen, usually the line profiles are well reproduced by the adopted model. The panels show the fit of the line profiles at the same locations, except for the bottom-right panels which are from distinct positions for each line, selected to represent regions where the line profiles are narrow.

\begin{figure}
    \centering
    \includegraphics[width=0.48\textwidth]{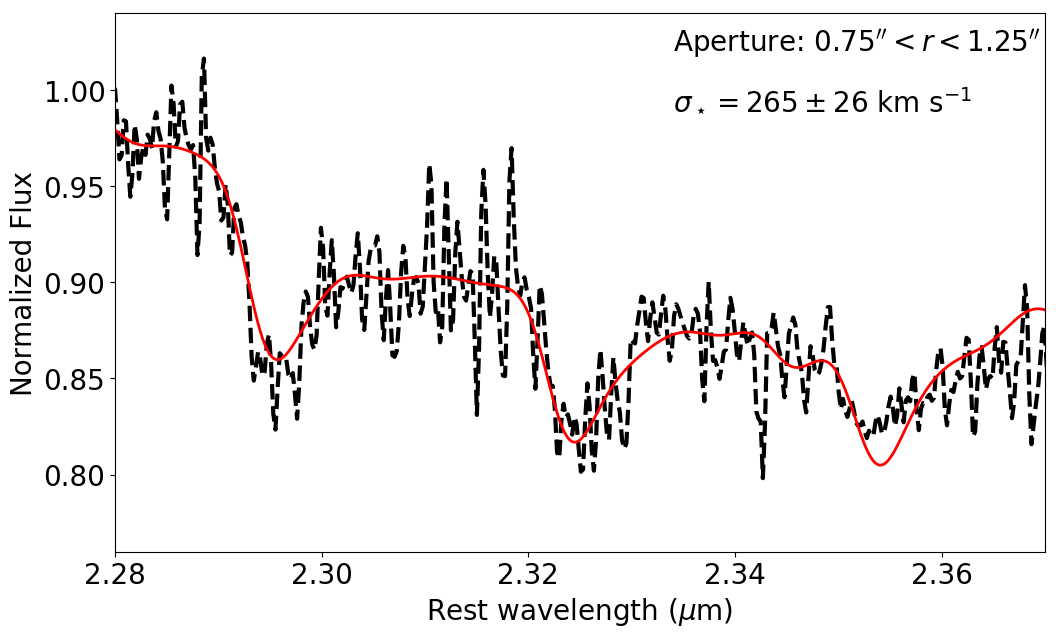}
    \caption{Observed (dashed black) and fitted (red) spectra in the CO absorption bands integrated within a circular ring with inner radius of 0\farcs75 and outer radius of 1\farcs25. }
    \label{stelfit}
\end{figure}

\begin{figure*}
    \centering
    \includegraphics[width=0.95\textwidth]{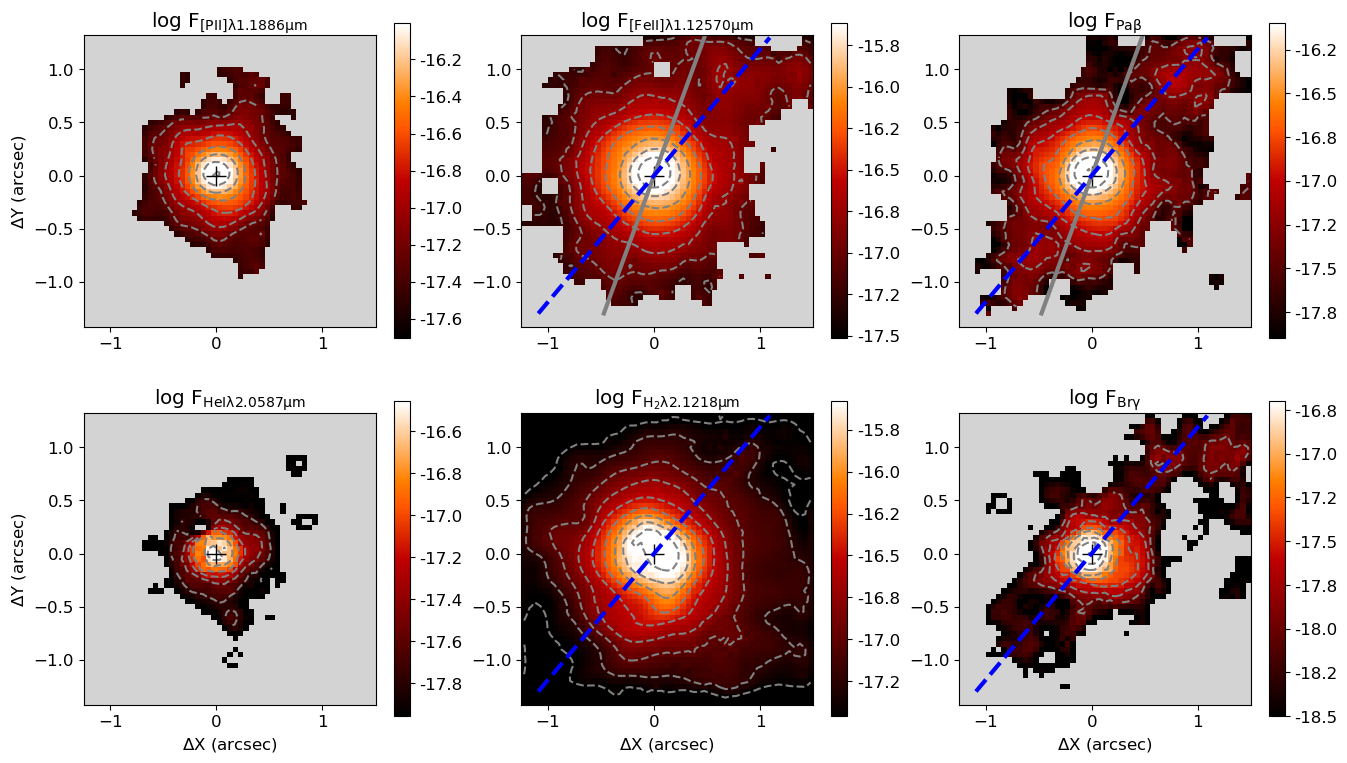}
    \caption{Flux maps for the [P\,{\sc ii}]$\lambda1.1886\,\mu$m, [Fe\,{\sc ii}]$\lambda1.2570\,\mu$m, Pa$\beta$, He\,{\sc i}$\lambda2.0587\,\mu$m, H$_2\lambda2.1218\,\mu$m and Br$\gamma$ emission lines. The central crosses mark the location of the peak of the continuum emission, the grey continuous line shows the orientation of the radio jet \citep{pedlar90}, the blue dashed line shows the orientation of the Pa$\beta$ emission and the color bars show the fluxes in logarithmic scale in units of erg\,s$^{-1}$cm$^{-2}$spx$^{-1}$. The gray regions represent masked locations, where the amplitude of the corresponding line profile is smaller than 3$\sigma$ of the adjacent continuum. North is up and east is to the left in all panels.}
    \label{fluxmaps}
\end{figure*}

\subsection{Stellar kinematics}

The CO absorption bandheads in the spectral range $\sim$2.29--2.40\,$\mu$m are usually prominent features in the spectra of nearby galaxies \citep[e.g.][]{rogerio06,mason15,rogerio19} and have been extensively used to measure the stellar kinematics \citep[e.g.][]{rogemar15,rogemar17}. However, in strong AGN the CO features can be diluted by the AGN continuum \citep{rogerio09,burtscher15,ms18}. This seems to be the case of NGC\,1275, where integrated spectra of the inner $r\leq150$\,pc do not show the CO features \citep{rogerio06,nifs13}. From our NIFS data, we are able to detect the CO absorption features in extranuclear regions, where we can therefore measure the stellar velocity and velocity dispersion. The signal-to-noise ratio (SNR) in the CO bands is not high enough to obtain reliable measurements spaxel by spaxel and thus, we use an integrated spectrum within a ring with inner radius of 0\farcs75 and outer radius of 1\farcs25.  For distances smaller than 0\farcs75, the CO absorptions are strongly diluted by the AGN continuum.

We measure the stellar line-of-sight velocity distribution by fitting the integrated spectrum within the spectral range $\sim$2.28--2.37\,$\mu$m (rest wavelengths). We use the penalized Pixel-Fitting ({\sc ppxf}) method \citep{cappelari04,cappelari17}, that finds the best fit to a galaxy spectrum by convolving  stellar spectra templates with a given velocity distribution, assumed to be Gaussian, as we fit only the two first moments. As spectral templates, we use the spectra of the Gemini library of late spectral type stars observed with the Gemini Near-Infrared Spectrograph (GNIRS) IFU and  NIFS \citep{winge09}, which includes stars with spectral types from F7 to M5.

Figure~\ref{stelfit} shows the integrated spectrum as a black dashed line and the model as a continuous red line.  The CO bandheads are clearly detected and the model reproduces well the observed spectrum. The best fit corresponds to a velocity of $V_S=5284\pm21\,{\rm km\,s^{-1}}$, corrected to the heliocentric  frame, and a velocity dispersion of $\sigma_\star=265\pm26\,{\rm km\,s^{-1}}$. The derived velocity is in agreement with measurements using optical emission lines \citep[5264$\pm$11 km\,s$^{-1}$; ][]{strauss92}. The value of $\sigma_\star$ derived here represents the first measurement based on the fitting of the CO bandheads. It is in agreement with those derived from Mg b lines at 5174\,\AA\  \citep[246$\pm$18 km\,s$^{-1}$; ][]{smith90}, Ca II$\lambda\lambda\lambda$8498,8542,8662 \citep[272$\pm$61 km\,s$^{-1}$; ][]{nelson95} and  using scaling relations between the stellar and gas velocity dispersion \citep[258.9$\pm$13.4 km\,s$^{-1}$; ][]{ho09}.

\section{Results}
In this section we present the two-dimensional maps produced by the methodology described in the previous section. In all maps, the gray regions correspond to masked locations where the amplitude of the corresponding emission-line is smaller than 3 times the standard deviation of the continuum next to the line. The north is to the top and east is to the left in all maps. The systemic velocity of the galaxy is subtracted  in all velocity maps, assumed to be the value derived from the fitting of the CO absorption bandheads.

\subsection{Emission-line flux distributions and ratios}

Figure~\ref{fluxmaps} shows the flux distributions of the [P\,{\sc ii}]$\lambda1.1886\,\mu$m, [Fe\,{\sc ii}]$\lambda1.2570\,\mu$m, Pa$\beta$, He\,{\sc i}$\lambda2.0587\,\mu$m, H$_2\lambda2.1218\,\mu$m and Br$\gamma$ emission lines, as obtained by direct integration of the observed line profiles. All emission lines present the peak emission at the nucleus of the galaxy and show spatially resolved emission. The  H$_2$ emission is the most extended, observed over the whole NIFS field of view, with the highest intensity levels  slightly more elongated to the south-west of the nucleus, in good agreement with the maps shown by \citet{nifs13}. The [Fe\,{\sc ii}]$\lambda1.2570\,\mu$m emission shows a similar morphology as that of the [Fe\,{\sc ii}]$\lambda1.6440\,\mu$m presented by \citet{nifs13}, showing the most extended features to up to 1\farcs7 north-west of the nucleus. The [P\,{\sc ii}] emission is extended to up to 1$^{\prime\prime}$ with a round morphology. The  He\,{\sc i} shows the most compact flux distribution, with emission seen mostly at distances smaller than 0\farcs5 from the nucleus. 

We detect extended emission in both H recombination lines, in contrast with the map presented by \citet{nifs13}, where the Br$\gamma$ emission is restricted to the inner 0\farcs3 radius. Both  Br$\gamma$ and Pa$\beta$ flux maps show a well defined linear structure along PA=140/320$^\circ$. These lines are consistent with being photo-ionized by the central AGN. Therefore, in this object the photo-ionization cone and the jet are aligned within 20$^\circ$. This is interesting because the jet and cone orientation are formed by processes on different scales  \citep{drouart12,bianchi12,marin16}.

\begin{figure*}
    \centering
    \includegraphics[width=0.65\textwidth]{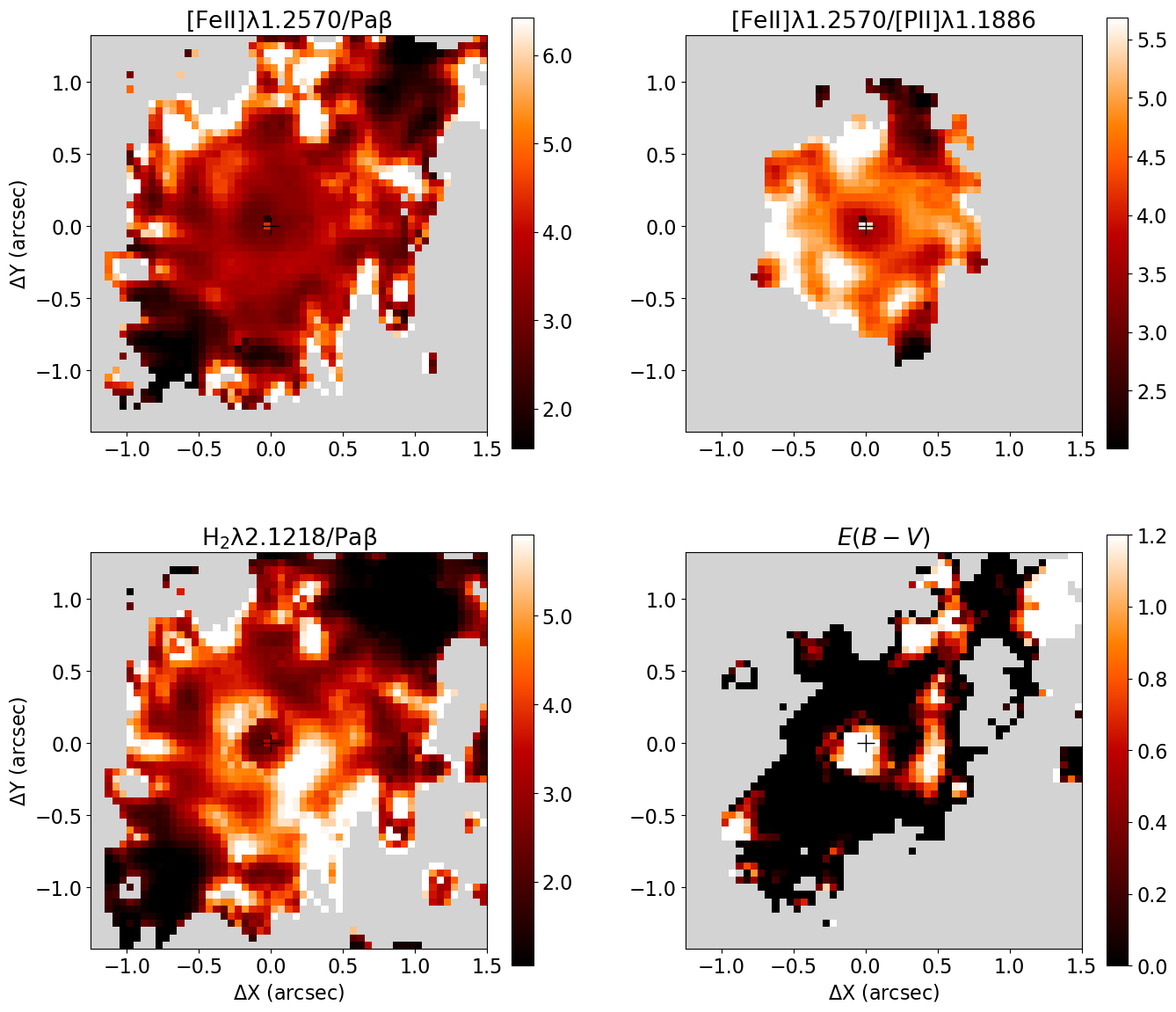}
    \caption{[Fe\,{\sc ii}]$\lambda1.2570\,\mu$m/Pa$\beta$, Fe\,{\sc ii}]$\lambda1.2570\,\mu$m/[P\,{\sc ii}]$\lambda1.1886\,\mu$m, H$_2\lambda2.1218\,\mu$m/Pa$\beta$ line ratio and $E(B-V)$ maps.}
    \label{ratiomaps}
\end{figure*}

\begin{figure*}
    \centering
    \includegraphics[width=0.85\textwidth]{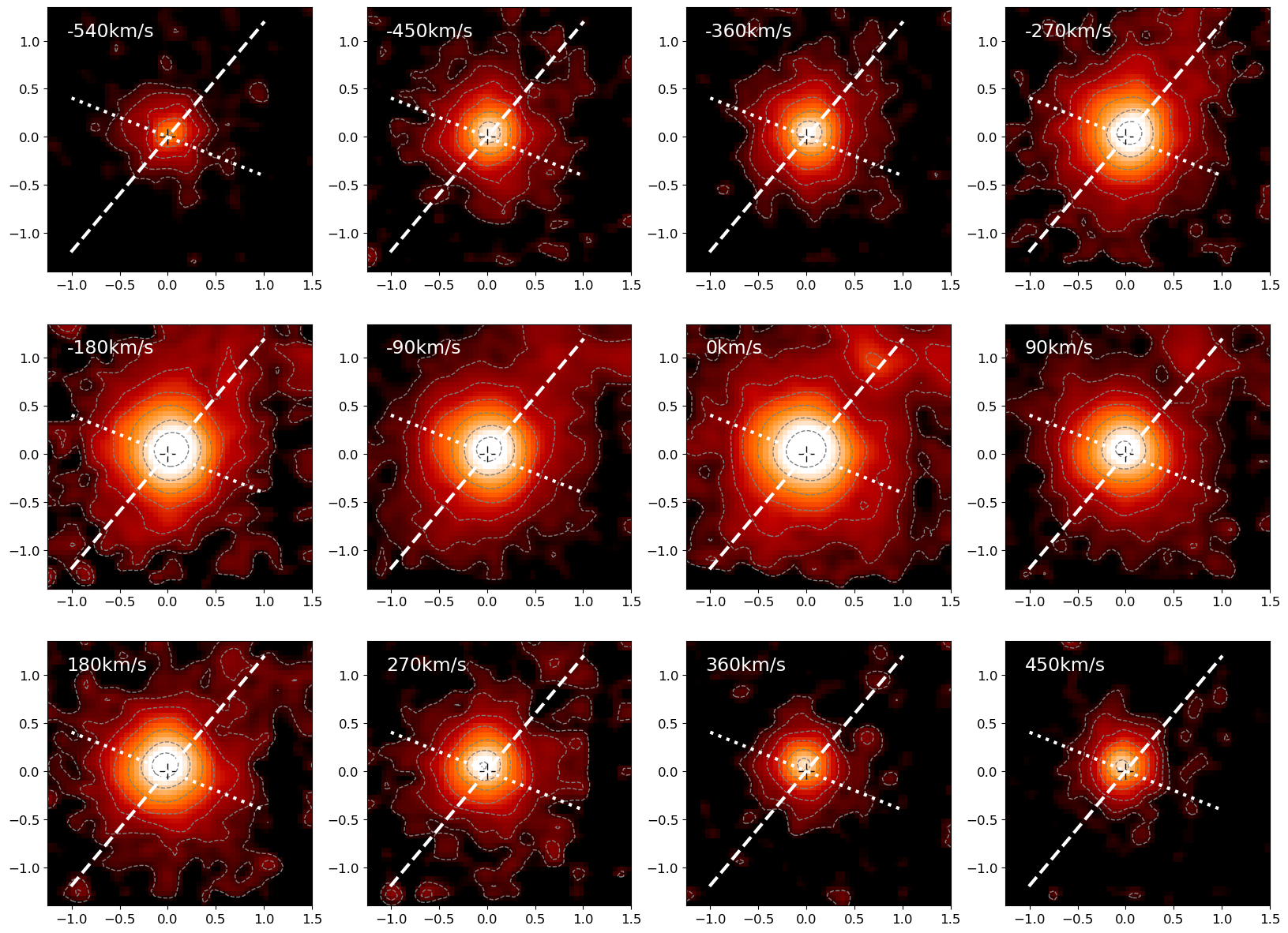}
    \caption{Velocity channel maps across the [Fe\,{\sc ii}]$\lambda1.2570\,\mu$m  emission-line profiles. The velocities are shown in the top-left corner of each panel and we use a velocity increment of 90\,km\,s$^{-1}$ from one channel to its subsequent. The dashed line shows the orientation of the AGN ionization structure and the dotted line represents the orientation of the compact gas disk. The central crosses mark the location of the nucleus of the galaxy. }
    \label{chamapFeII}
\end{figure*}

\begin{figure*}
    \centering
    \includegraphics[width=0.85\textwidth]{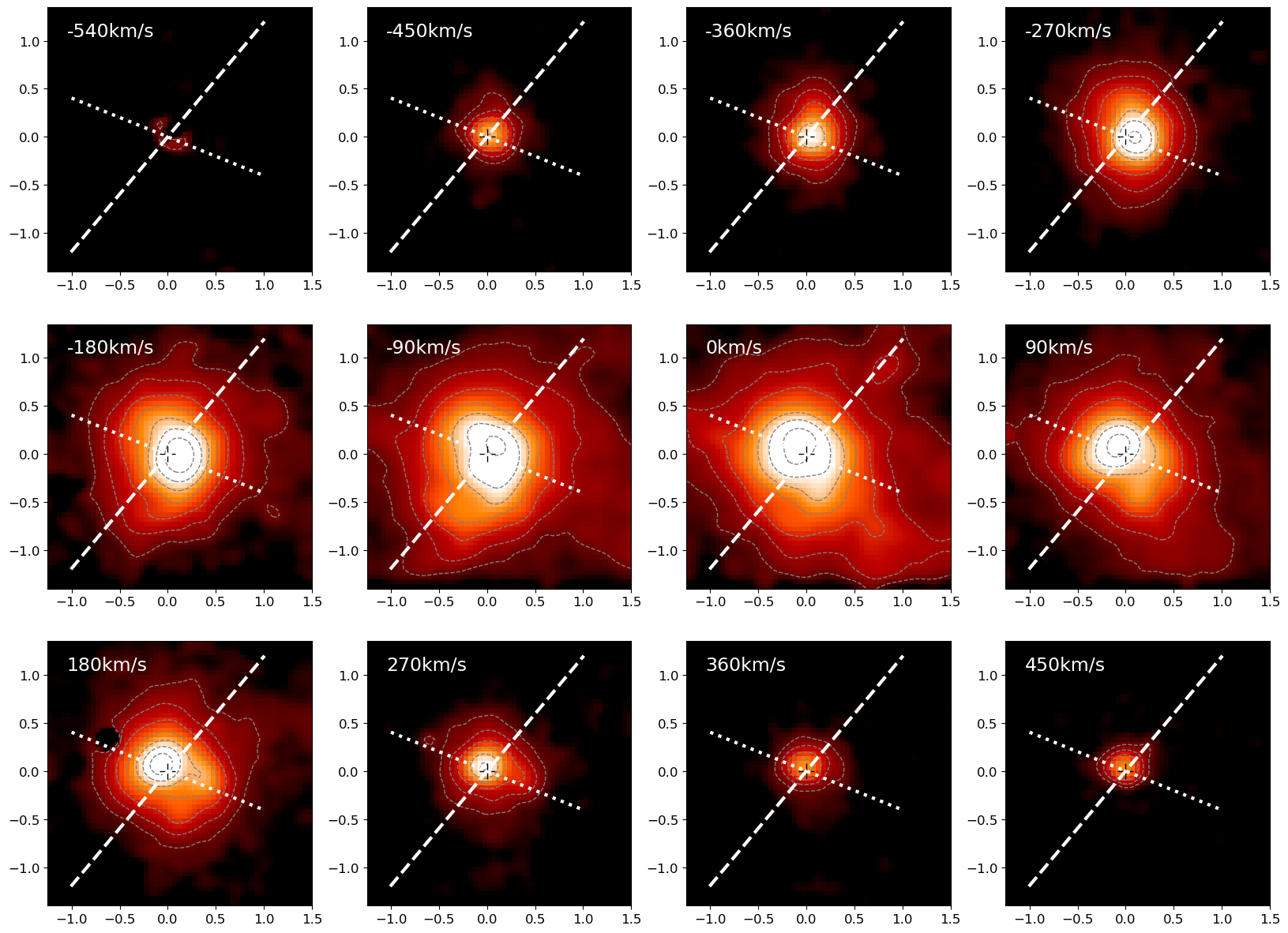}
    \caption{Same as Fig.\,\ref{chamapFeII} for the  H$_2\lambda2.1218\,\mu$m emission line.}
    \label{chamapH2}
\end{figure*}

Figure~\ref{ratiomaps} presents the [Fe\,{\sc ii}]$\lambda1.2570\,\mu$m/Pa$\beta$, [Fe\,{\sc ii}]$\lambda1.2570\,\mu$m/[P\,{\sc ii}]$\lambda1.1886\,\mu$m, H$_2\lambda2.1218\,\mu$m/Pa$\beta$ line ratio and color excess  maps. We estimate the color excess by:
\begin{equation}
 E(B-V)=4.74\,{\rm log}\left(\frac{5.88}{F_{Pa\beta}/F_{Br\gamma}}\right),
\end{equation}
where $F_{Pa\beta}$ and $F_{Br\gamma}$ are the fluxes of Pa$\beta$ and Br$\gamma$ emission lines, respectively. We adopt the theoretical ratio between Pa$\beta$ and Br$\gamma$ of $5.88$, corresponding to case B at the low-density limit \citep{oster06} for an electron temperature of $T_e=10^4$\,K and we use the reddening  law of \citet{ccm89}.

The [Fe\,{\sc ii}]$\lambda1.2570\,\mu$m/Pa$\beta$ and  H$_2\lambda2.1218\,\mu$m/Pa$\beta$ can be used to investigate the origin of the [Fe\,{\sc ii}] and H$_2$ emission. Starburst galaxies and H\,{\sc ii} regions present [Fe\,{\sc ii}]$\lambda1.2570\,\mu$m/Pa$\beta<0.6$ and H$_2\lambda2.1218\,\mu$m/Pa$\beta<0.07$. For emission photo-ionized by an AGN typical values are:  $0.6<$[Fe\,{\sc ii}]$\lambda1.2570\,\mu$m/Pa$\beta<2.0$ $0.07<$H$_2\lambda2.1218\,\mu$m/Pa$\beta<1.0$, while higher ratios may indicate the contribution by shocks to the H$_2$ and [Fe\,{\sc ii}] emission \citep{reunanen02,ardila04,ardila05,rogemarMrk1066exc,rogerio13,colina15,lamperti17,rogemar_he210}. As seen in Fig.~\ref{ratiomaps}, both line ratios show a wide range of values. The smallest values of [Fe\,{\sc ii}]$\lambda1.2570\,\mu$m/Pa$\beta$ = 1--2 and H$_2\lambda2.1218\,\mu$m/Pa$\beta$=0.5--1.0 are seen only in two small regions (0\farcs5 diameter each) along PA=140/320$^\circ$ -- at $\sim$1\farcs3 northwest and  $\sim$1\farcs1 southeast of the nucleus, coincident with the orientation of the linear emission seen in Br$\gamma$ and Pa$\beta$. This indicates that the AGN photo-ionization 
 plays some role in observed line emission from these locations and that the linear structure seen in the H recombination lines is tracing the AGN ionization axis. However, in most locations both line ratios show values much higher than commonly observed in AGN, in particular from locations away from the ionization axis. 
 
 The [Fe\,{\sc ii}]$\lambda1.2570\,\mu$m/Pa$\beta$  map shows values of up to 6, while H$_2\lambda2.1218\,\mu$m/Pa$\beta$ presents values higher than 5. Considering that most of the gray regions in these maps are due to the non-detection of Pa$\beta$ -- as the H$_2$ and [Fe\,{\sc ii}] emission is clearly more extended in the maps of Fig.~\ref{fluxmaps} -- the highest ratios can be considered lower limits for the gray regions. This indicates that shocks play an important role not only in the production of the H$_2$ emission, as pointed out in \citet{nifs13}, but also in the [Fe\,{\sc ii}] emission at least in locations away from the ionization cone.    

Another line ratio that can be used as a tracer of shocks is  [Fe\,{\sc ii}]$\lambda1.2570\,\mu$m/[P\,{\sc ii}]$\lambda1.1886\,\mu$m, as the P$^+$ and Fe$^+$ have similar ionization potential, radiative recombination coefficients, and  ionization temperatures \citep{oliva01,sb09,diniz19}. For H\,{\sc ii} regions this line ratio is $\sim$2, while if shocks are present the [Fe\,{\sc ii}] is enhanced as the shocks release the Fe from the dust grains \citep[e.g.][]{oliva01}. Values of [Fe\,{\sc ii}]$\lambda1.2570\,\mu$m/[P\,{\sc ii}]$\lambda1.1886\,\mu$m$>2$ cannot be reproduced by photo-ionization models \citep[e.g.,][]{rogerio19} and indicate that shocks contribute to the origin of the [Fe\,{\sc ii}] emission. For NGC\,1275, this line ratio shows values larger than 2 in most locations and reaching [Fe\,{\sc ii}]$\lambda1.2570\,\mu$m/[P\,{\sc ii}]$\lambda1.1886\,\mu$m$\approx6$ at distances of 0\farcs5 from the nucleus. As [Fe\,{\sc ii}] shows more extended emission than the [P\,{\sc ii}], the value of 6 can be considered a lower limit for the [Fe\,{\sc ii}]$\lambda1.2570\,\mu$m/[P\,{\sc ii}]$\lambda1.1886\,\mu$m ratio in locations masked out due to the non-detection of [P\,{\sc ii}] emission.  This result further supports the importance of shocks for the  [Fe\,{\sc ii}] emission in NGC\,1275.

The $E(B-V)$ map (Fig.~\ref{ratiomaps}) shows values close to zero at most locations, in agreement with previous results that show that the extinction in the central region of NGC\,1275 is low \citep[e.g.,][]{kent79,fabian84,johnstone95}. The highest values of up $1.0$ are seen at the nucleus and to the northwest of it,  the same location where the H$\alpha$ filaments are more concentrated and seen closer to the nucleus \citep[e.g.][]{fabian08}. 

\subsection{Gas kinematics}

A way to map the gas kinematics is by constructing velocity channel maps along the emission line profiles, which allows a better mapping of the emission at the  wings (higher velocities). We show such maps for the [Fe\,{\sc ii}] and H$_2$ emission lines in Figures~\ref{chamapFeII} and \ref{chamapH2}. We do not show channel maps for other emission lines, 
as the emission is weaker and the channel maps noisier. But it is possible to see that the Pa$\beta$ channel maps are similar to those of [Fe\,{\sc ii}].  
Each panel shows the fluxes integrated within a velocity bin of 90\,km\,s$^{-1}$, centred at the velocity shown in the top-left corner, relative to the systemic velocity of the galaxy. 
Both lines show emission in velocities up to 500\,km\,s$^{-1}$, seen both in blueshifts (negative velocities) and redshifts (positive velocities).

For velocities in the range from $-$180 to 180~km\,s$^{-1}$, the peak of emission in the inner 0\farcs3 moves from southwest in blueshifted channels to northeast in redshifted channels (clearly seen in the H$_2$ channel maps). This structure is tracing the emission of the compact disk seen in both ionized and molecular gas with major axis oriented along $PA=68^\circ$ \citep{nifs13,nagai19}. Besides this structure, in low velocity channels it can be seen that the H$_2$ and [Fe\,{\sc ii}] present distinct flux distributions. While the [Fe\,{\sc ii}] presents most of the emission approximately along the orientation of the ionization axis (PA=140/320$^\circ$), the H$_2$ emission is seen mainly in the perpendicular direction. At the highest velocities, both lines show the peak of emission at the nucleus and extended emission is seen to at least 0\farcs5. We do not take into account the effect of the beam smearing in the determination of the extensions, but given that sizes of most structures are much larger than the angular resolution of our data ($\sim$0\farcs2) this effect is negligible.

\begin{figure*}
    \centering
    \includegraphics[width=0.95\textwidth]{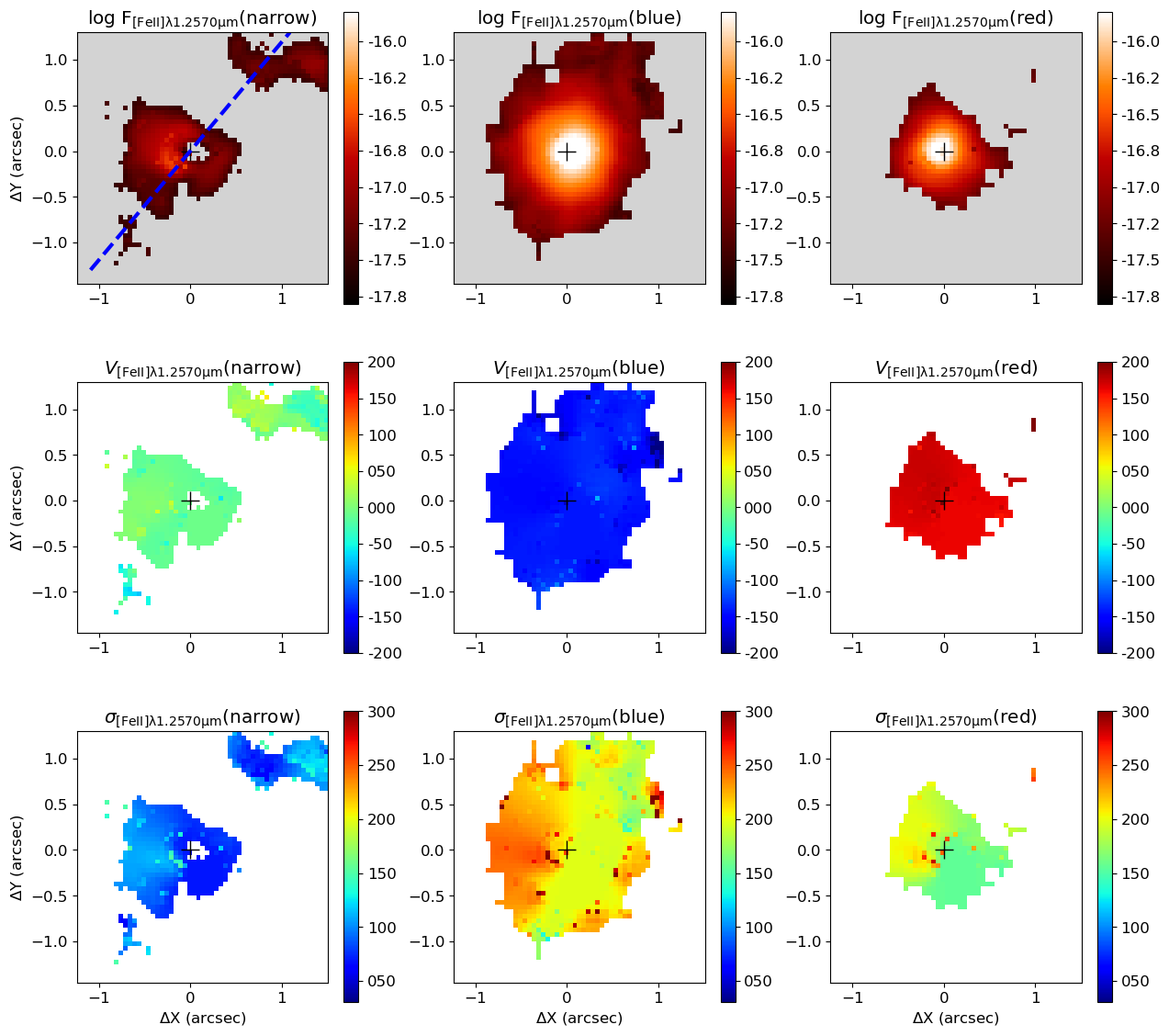}
    \caption{Flux, velocity and velocity dispersion maps from multi-Gaussian components fits of the [Fe\,{\sc ii}]$\lambda1.2570\,\mu$m emission-line profile. The dashed line shows the orientation of the collimated Pa$\beta$ emission.}
    \label{FeIImg}
\end{figure*}

We also fit the emission-line profiles of [Fe\,{\sc ii}]$\lambda1.2570\,\mu$m, Pa$\beta$  and H$_2\lambda2.1218\,\mu$m by multi-Gaussian components, as described in Sec.~\ref{ELfits}.
The resulting maps for flux, velocity and velocity dispersion are shown in  Figs.\ref{FeIImg}, \ref{Pabmg} and \ref{H2mg}. All lines present three kinematic components: a narrow ($\sigma\lesssim150\,$km\,s$^{-1}$) and two broad components ($\sigma\gtrsim150\,$km\,s$^{-1}$) -- see also Fig.~\ref{linefits}. In addition, Pa$\beta$ presents a very broad component with spatially unresolved flux distribution produced by the emission of the BLR. The narrow component is observed at velocities close to the systemic velocity of the galaxy, while the broad components are observed  blueshifted and redshifted by $150-200$\,km\,s$^{-1}$.  The peak of the blueshifted component is slightly shifted to the southwest relative the nucleus and the redshifted components peaks slightly northeast of it, consistent with the compact rotating disk seen in the inner region of the galaxy.

\begin{figure*}
    \centering
    \includegraphics[width=0.95\textwidth]{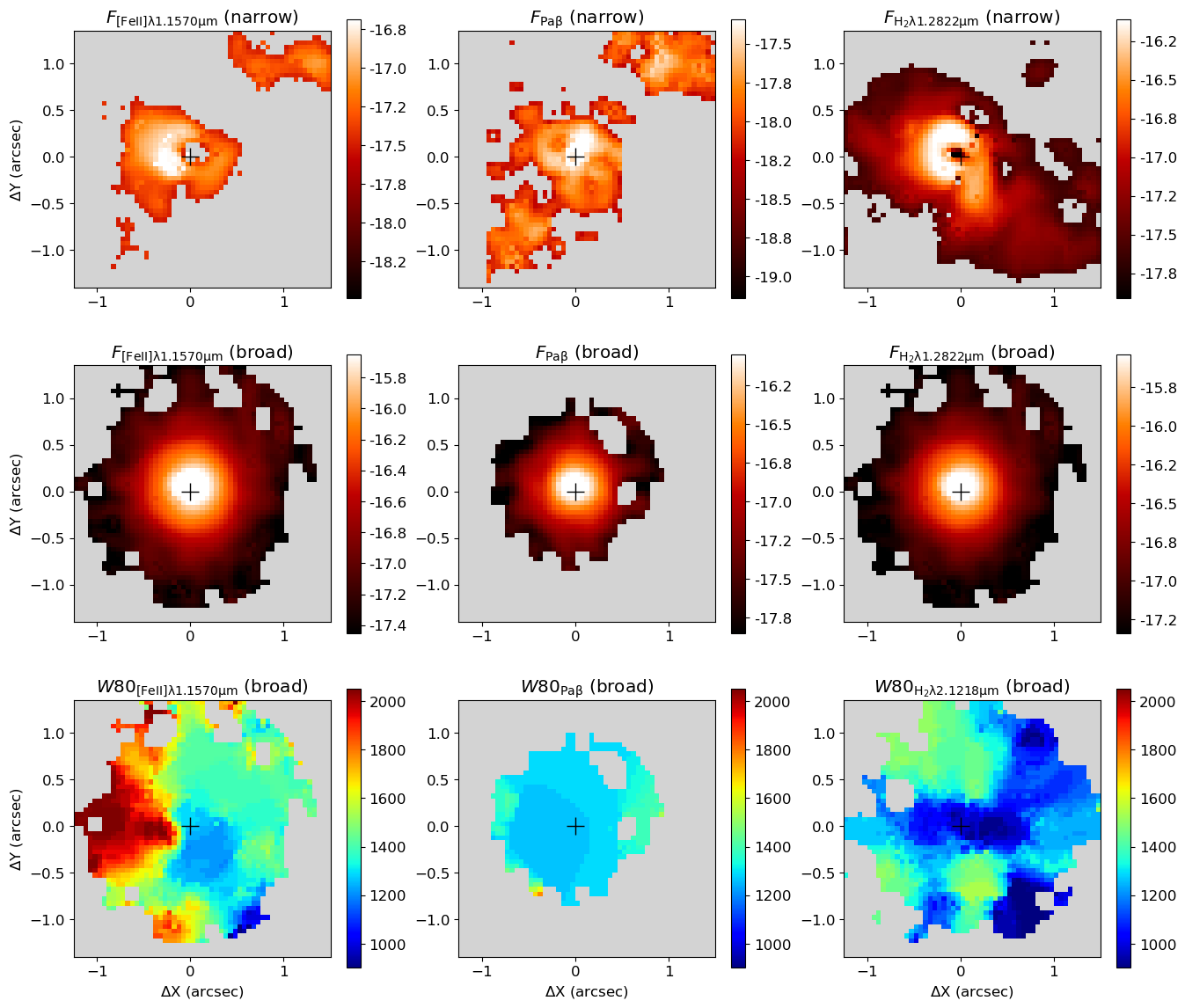}
    \caption{Flux maps of the narrow  (top) and broad (middle) line components, and $W_{80}$ constructed after subtracting the narrow line component.  The fluxes of the broad component are constructed by summing the fluxes of two broad line components (Fig.~\ref{linefits}). Gray regions are masked locations where the amplitude of the fitted Gaussian is smaller than 3 times the standard deviation of the adjacent continuum.}
    \label{maps_subnarrow}
\end{figure*}

In Figure \ref{maps_subnarrow} we present the flux maps for the narrow and broad components of  [Fe\,{\sc ii}]$\lambda1.2570\,\mu$m, Pa$\beta$  and H$_2\lambda2.1218\,\mu$m  emission lines. The maps for the broad component are constructed by summing the fluxes of the blueshifted and redshifted broad components. The flux distributions of the narrow components of [Fe\,{\sc ii}]$\lambda1.2570\,\mu$m and Pa$\beta$ show a linear structure along  PA=140/320$^\circ$, while the H$_2\lambda2.1218\,\mu$m presents narrow emission mostly perpendicular to the extended emission seen in ionized gas.  As the narrow component presents low velocity dispersion and low velocities, its origin may be emission from the galaxy disk. The [Fe\,{\sc ii}]$\lambda1.2570\,\mu$m and Pa$\beta$ seem to be tracing emission of gas of the disk ionized by the AGN radiation, while most of the H$_2$ emission originates away from the ionization axis, as seen in other nearby Seyfert galaxies \citep{sb09,rogemar_n1068,may17}. The broad component of all emission lines show similar behaviours with round flux distributions and centrally peaked emission. The [Fe\,{\sc ii}] and H$_2$ emission are observed over the inner 1\farcs5, while the Pa$\beta$ shows emission only in the inner $\sim$1\farcs0.

The bottom panels of Fig.\,\ref{maps_subnarrow} show the $W_{80}$ maps for the broad emission of [Fe\,{\sc ii}]$\lambda$1.2570\,$\mu$m, Pa$\beta$ and H$_2\lambda$2.1218\,$\mu$m. We compute the $W_{80}$ values from the modeled spectra, after the subtraction of the contribution of the narrow component from the galaxy disk. All maps show $W_{80}$  values larger than 1\,000 km\,s$^{-1}$ in all regions. The smallest values are seen for H$_2$, followed by Pa$\beta$ and the highest values of up to 2\,000 km\,s$^{-1}$ are seen for the [Fe\,{\sc ii}], mostly to the east of the nucleus.

\begin{figure*}
    \centering
    \includegraphics[width=0.4\textwidth]{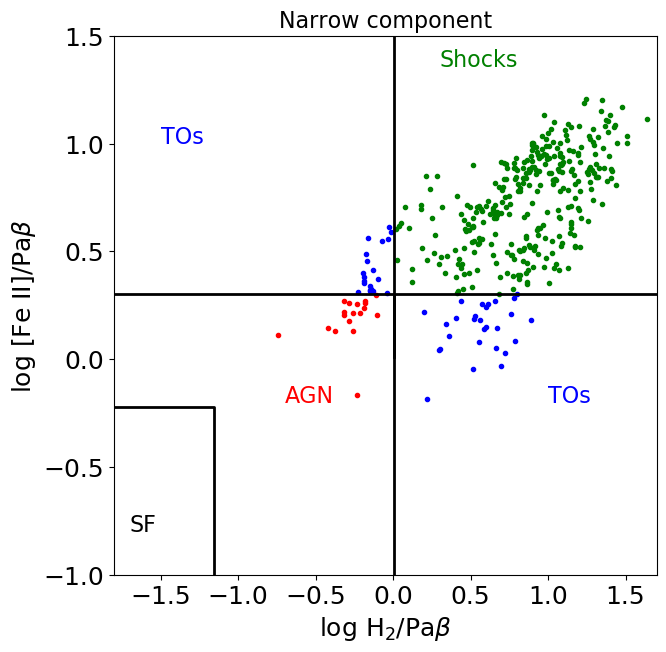}
    \includegraphics[width=0.4\textwidth]{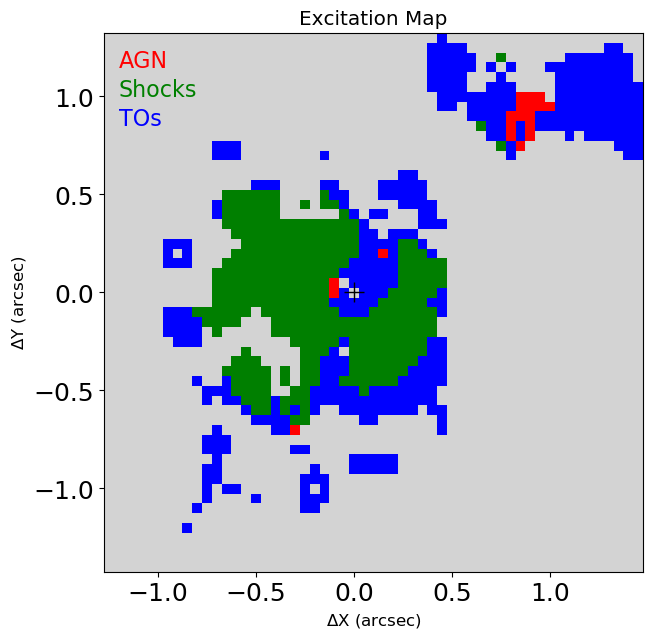}
    \includegraphics[width=0.4\textwidth]{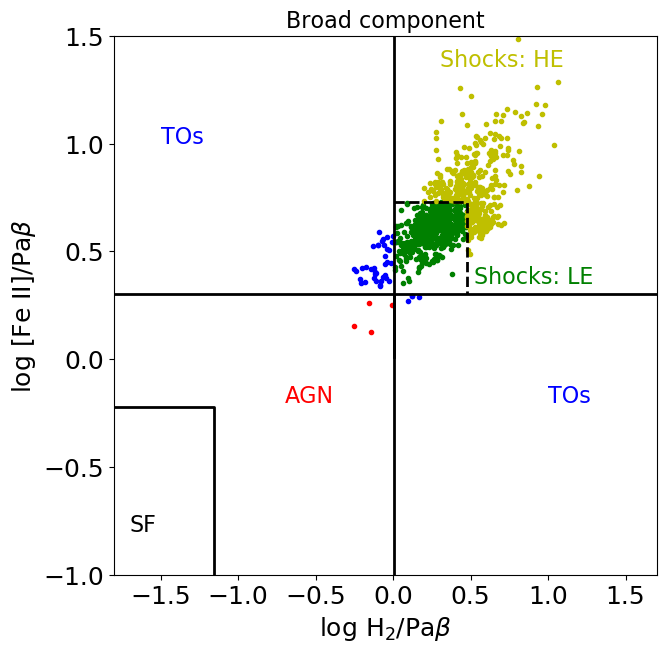}
    \includegraphics[width=0.4\textwidth]{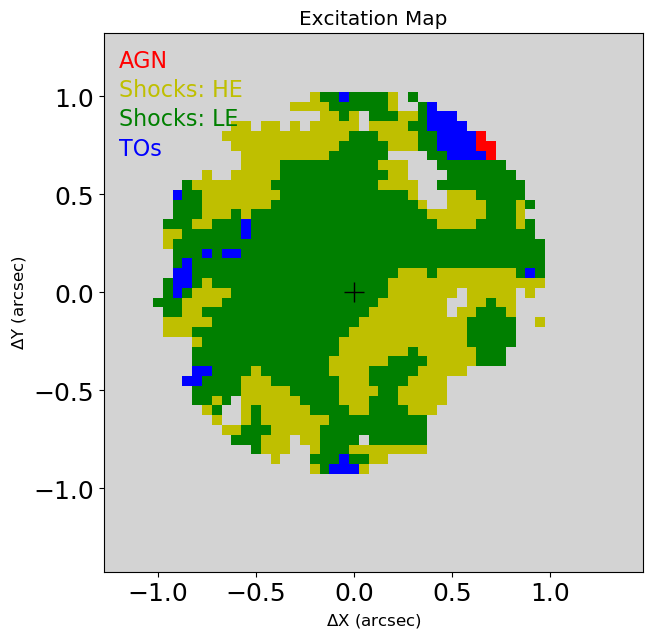}
    \caption{H$_2\lambda$2.1218$\,\mu$m/Pa$\beta$ vs. [Fe\,{\sc ii}]$\lambda1.2570\,\mu$m/Pa$\beta$ line-ratio diagnostic diagrams for the narrow (top) and broad (bottom) components using the fluxes shown in Fig.\,\ref{maps_subnarrow}. The black lines separate regions where the line emission is produced star formation (SF), AGN, shocks and transition objects (TOs) as defined \citet{rogerio13} using a large sample of galaxies observed with long-slit spectroscopy.  The dashed lines shown in the diagram for the outflow component separate regions of high excitation (log  [Fe\,{\sc ii}]$\lambda1.2570\,\mu$m/Pa$\beta$ > 0.73 and  H$_2\lambda2.1218\,\mu$m/Pa$\beta$ > 0.47) and low excitation by shocks. We adopt the theoretical ratio Pa$\beta$/Br$\gamma=5.88$ \citep{oster06} to convert the Br$\gamma$  (presented in \citet{rogerio13}) to Pa$\beta$ flux and draw the separation lines.} 
    \label{diagnostic}
\end{figure*}

In Figure\,\ref{diagnostic} we present the H$_2\lambda$2.1218$\,\mu$m/Pa$\beta$ vs. [Fe\,{\sc ii}]$\lambda1.2570\,\mu$m/Pa$\beta$ line-ratio diagnostic diagrams for the narrow and broad components obtained using the fluxes shown in Fig.~\ref{maps_subnarrow}. The right panels of Fig.\,\ref{diagnostic} show the color coded excitation maps, identifying the spatial location of each region of the diagram. We adopt the separation lines from \citet{rogerio13}, based on long-slit spectroscopy of 67 emission-line galaxies. 
For the narrow component one or both line ratios are consistent with an AGN origin along the AGN ionization axis, while shocks seem to dominate the gas excitation in regions perpendicular to the ionization cone.  

The line ratios for the broad component are consistent with emission produced by shocks in all locations. To separate between regions of high and low gas excitation, we compute the 75th percentile ($P_{75}$) of the H$_2\lambda$2.1218$\,\mu$m/Pa$\beta$ and [Fe\,{\sc ii}]$\lambda1.2570\,\mu$m/Pa$\beta$ line ratios of points in the shock region of the diagram, and obtain  log $P_{75}$(H$_2\lambda$2.1218$\,\mu$m/Pa$\beta$)=0.47 and  $P_{75}$([Fe\,{\sc ii}]$\lambda1.2570\,\mu$m/Pa$\beta$)=0.73. The limits delineated by these values are shown as a dashed line the in diagnostic diagram for the broad component.   

\section{Discussion}

\subsection{The supermassive black hole mass}

 The mass of the black hole of NGC\,1275 is $M_{\rm SMBH}\sim1\times10^9$\,M$_\odot$, as obtained by modeling the H$_2$ \citep{nifs13} and CO(2--1) kinematics \citep{nagai19} and adopting a disk inclination of $i=45^\circ\pm10^\circ$. This dynamical mass is about two orders of magnitude larger than that obtained from measurements  
 of the very broad components (from the BLR) of near-infrared recombination lines and X-ray luminosity \citep[$M_{\rm SMBH}=(2.9\pm0.4)\times10^7$\,M$_\odot$][]{onori17}.

 \citet{onori17} measure the near-infrared line widths using the spectrum from \citet{rogerio06} and obtain   FWHM$_{\rm Pa\beta \,BLR}=2824^{+98}_{-85}$ km\,s$^{-1}$ for Pa$\beta$ and FWHM$_{\rm He I\,BLR}=2547^{+20}_{-24}$ km\,s$^{-1}$ for He\,{\sc i}$\lambda$1.083\,$\mu$m. Using a spectrum obtained by integrating the NIFS J-band datacube within an aperture of 0\farcs25 radius centred at the nucleus, we measure FWHM$_{\rm Pa\beta\, BLR}=5825\pm280$ km\,s$^{-1}$ for the very broad component of Pa$\beta$, which is about two times the value of \citet{onori17}. The low values measured by \citet{onori17} can be due to the fact that the very broad components of He\,{\sc i}$\lambda$1.083\,$\mu$m and Pa$\beta$ are weak and strongly blended with the narrow line emission in the spectrum of \citet{rogerio06}. The very broad component of the Pa$\alpha$ emission line is strong in their spectrum and we measure FWHM$_{\rm Pa\alpha\,BLR}=5607\pm350$ km\,s$^{-1}$, which is consistent with the value measured for Pa$\beta$ using our NIFS data. Our value of FWHM$_{\rm Pa\beta\,BLR}$ is also in good agreement with recent results for the broad component of H\,{\sc i} recombination lines \citep{punsly18}. Using our measurement of  FWHM$_{\rm Pa\beta\,BLR}$, together with the X-ray luminosity of the point source and equation 1 of \citet{onori17}, we obtain $M_{\rm SMBH}=(1.37\pm0.13)\times10^8$\,M$_\odot$. The mass of the supermassive black hole can also be estimated using the width of the Pa$\beta$ line and near-infrared luminosity \citep{landt13}. We measure $\log(\frac{\nu L_\nu}{\rm erg s^{-1}})=43.2$ at 1.15\,$\mu$m integrated within 0\farcs25 radius from the nucleus. Using this value, together with  FWHM$_{\rm Pa\beta\,BLR}=5825\pm280$ km\,s$^{-1}$, estimated above, and equation 2 from \citet{landt13} we estimate   $M_{\rm SMBH}=1.8^{+0.9}_{-0.6}\times10^8$\,M$_\odot$, which is consistent with the value obtained using the X-ray luminosity, but about one order of magnitude smaller than the dynamical determinations. 
 
 Using our measurement of the stellar velocity dispersion -- $\sigma_\star=265\pm26\,{\rm km\,s^{-1}}$ -- and the $M_{\rm SMBH}-\sigma_\star$ calibration of \citet{kormendy13}, we obtain $M_{\rm SMBH}=1.1^{+0.9}_{-0.5}\times10^9$\,M$_\odot$, which is in good agreement with the dynamical determinations \citep{nifs13,nagai19}. The origin of discrepancy between the values of $M_{\rm SMBH}$ measured using  scaling relations with line widths and luminosities and those obtained from the  $M_{\rm SMBH}- \sigma_\star$ relation and gas dynamics is unclear and beyond the scope of this work, but could be related to the uncertainties in calibration of the distinct methods.  
 
\subsection{The compact disk}

 NGC\,1275 is one the most studied Seyfert galaxies. These studies include results from near-IR integral field spectroscopy that can be directly compared with our new maps for the inner $\sim$500\,pc of the galaxy. The most prominent kinematic feature, revealed by these studies is a compact rotating disk in the inner 100 pc of the galaxy, seen both in ionized and hot molecular gas \citep{wilman05,nifs13}. The cold counterpart of this disk is detected by recent observations of CO(2--1) line using the ALMA \citep{nagai19}. The major axis of the disk is oriented along PA=68$^\circ$ and the line-of-sight velocity amplitude is $\sim$150 km\,s$^{-1}$  and a mass of gas of $\sim10^8$ M$_\odot$ \citep{nifs13,nagai19}.
 
  As in previous works, the kinematic component attributed to a compact disk is  observed in our data (e.g. Fig.~\ref{chamapH2}), which shows that the peak of emission in the inner 0\farcs3 moves from southwest  to northeast for velocities in the range from $-$180 to 180~km\,s$^{-1}$). By fitting the emission-line profiles by a single Gaussian components, we obtain velocity fields very similar to those presented by \citet{nifs13}, and thus we do not present them in this paper.  Alternatively, the kinematic component previously identified as a compact disk could be a misinterpretation of the results due to a simplified approach in the modeling of the shape of the observed line, i.e. a single Gaussian, which is clearly not the best model for the observed near-IR line profiles.  In this case, the interpretation of the multi-Gaussian components would be straightforward, with the narrow component due to an almost face on disk partially illuminated by the AGN radiation and the two broad components due to a centrally driven outflow.

\subsection{AGN ionization structure}
 
 Besides confirming previous results, our new data set and analysis reveal new kinematic and morphological structures in the inner region of NGC\,1275, leading to a breakthrough in our understanding of the nature of its central AGN. Here, we focus our discussion on these new findings, rather than on the previous known results. 
 We find extended narrow line ($\sigma<150$\,km\,s$^{-1}$) emission from the ionized gas along PA=140/320$^\circ$, which seems to be tracing the AGN ionization axis and its surroundings (Fig. \ref{maps_subnarrow}). Indeed, the H$_2\lambda2.1218\,\mu$m/Pa$\beta$ vs.  [Fe\,{\sc ii}]$\lambda1.2570\,\mu$m/Pa$\beta$ diagnostic diagram (Fig.~\ref{diagnostic}) for the narrow component suggest some contribution of the AGN radiation in the production of the observed line emission. Along the major axis of the disk (NW--SE),  higher values are observed for both ratios with [Fe\,{\sc ii}]$\lambda1.2570\,\mu$m/Pa$\beta$ > 2 and H$_2\lambda2.1218\,\mu$m/Pa$\beta$>1, indicating some contribution of shocks in the disk. However, as the overlap region between locations of H$_2$, [Fe\,{\sc ii}] and Pa$\beta$ emission is small, we have to be cautious with conclusions based on these line ratios for the narrow component. The structure of this extended emission resemble an ionization cone with an opening angle of $\sim$50$^\circ$, with an axis tilted by 20$^\circ$ relative to the orientation of the radio jet \citep{pedlar90} and is  approximately orthogonal to the orientation of the compact disk, suggesting that the inner parts of the disk are responsible for obscuration of the AGN radiation. 
 
 We find that the radio jet and ionization axes are almost perpendicular to the compact disk. Whether or not there is an actual physical connection between the disk (and the dusty torus) and the jet is still an open question.  In a steady growth of a black hole via accretion along a preferred plane, the orientation of the radio jet is expected to be perpendicular to that of the disk \citep{drouart12,bianchi12,marin16}. Although we find that the jet and the ionization structure present similar projected orientation and are perpendicular to the disk, we do not find any clear evidence of jet-cloud interaction in the ionized and molecular gas distribution and kinematics. Indeed, the orientations of the AGN ionization axis and radio jet are very difficult to measure and uncertain.

 The H$_2$ presents narrow-line emission (top-right panel of Fig. \ref{maps_subnarrow}), mostly perpendicularly to the ionization axis and along the same orientation of the major axis of the compact disk (PA=68$^ \circ$). The narrow component of all emission lines is observed at velocities close to the systemic velocity of the galaxy. A possible interpretation for this kinematic structure is that it is due to the outer regions of the compact disk. To account for the small velocities, the disk at larger distances must be slightly warped relative to the orientation of the compact disk. \citet{fujita17} find that the sub-pc scale radio jet is inclined by 65$^\circ$ relative to the line-of-sight and if the compact disk is perpendicular to this structure ($i=25^\circ$, relative to the plane of the sky), a small warping in the outer regions could account for the low line-of-sight velocities. In addition, precession of the radio jet in NGC\,1275 has been suggested as the origin of its large scale X-ray cavities \citep[e.g.,][]{falceta-goncalves10}, and thus a misalignment between the inner and outer discs seems to be very likely.
 The fact that the H$_2$ and ionized gas emission are seen mostly perpendicular to each other may be due to dissociation of the H$_2$ molecule within the ionization cone by the AGN radiation field, as seen in other well studied nearby Seyfert galaxies  \citep[e.g.,][]{sb09,rogemar_n1068,shimizu19}. 
 
 \subsection{Shocks and outflows}
 
 Previous studies of NGC\,1275 indicate that the hot H$_2$ emission originates from thermal excitation caused by shocks \citep{krabbe00,ardila05,wilman05,nifs13}, which is consistent with our data.  \citet{nifs13} find that the origin of the [Fe\,{\sc ii}] line emission is consistent with X-ray heating of the gas with an electron density of $\sim$4\,000 cm$^{-3}$, but based only on  H and K-band emission lines, they could not rule out the contribution of fast shocks. The J band data allowed us to map the [Fe\,{\sc ii}]$\lambda1.2570\,\mu$m/[P\,{\sc ii}]$\lambda1.1886\,\mu$m and [Fe\,{\sc ii}]$\lambda1.2570\,\mu$m/Pa$\beta$ line ratios, and their high values indicate that shocks indeed play an important role in the production of the [Fe\,{\sc ii}] emission in NGC\,1275 \citep[e.g.,][]{oliva01,reunanen02,ardila04,colina15}. In particular, the [Fe\,{\sc ii}]$\lambda1.2570\,\mu$m/[P\,{\sc ii}]$\lambda1.1886\,\mu$m ratio is very sensitive to shocks as the ions have similar ionization potentials and recombination coefficients, and both lines have similar excitation temperatures. In objects were the lines are produced by photo-ionization, the  Fe is locked into dust grains and the [Fe\,{\sc ii}]$\lambda1.2570\,\mu$m/[P\,{\sc ii}]$\lambda1.1886\,\mu$m$\approx2$, while if shocks are present they release the Fe from the grains, enhancing its abundance and line intensity \citep{oliva01,sb09,rogemar_n1068}. In supernova remnants, where shocks dominates the [Fe\,{\sc ii}] excitation, this line ratio can be as high as 20 \citep{oliva01}. In NGC\,1275, the [Fe\,{\sc ii}]$\lambda1.2570\,\mu$m/[P\,{\sc ii}]$\lambda1.1886\,\mu$m map (Fig.~\ref{ratiomaps}) presents values larger than 2 in all locations, indicating that shocks contribute to the observed [Fe\,{\sc ii}] emission, but the ratios are never as high as 20, suggesting that shocks are not the only excitation mechanism. Considering only the fluxes of the broad component, we find that   H$_2\lambda2.1218\,\mu$m/Pa$\beta$ and  [Fe\,{\sc ii}]$\lambda1.2570\,\mu$m/Pa$\beta$ line ratios are consistent with emission produced by shocks in all regions. As shown in the excitation map of Fig.~\ref{diagnostic}, the gas of higher excitation is observed mainly perpendicular to the ionization cone. This can be explained if the gas is affected both by shocks and by photo-ionization, than photo-ionization dominates line excitation as shocks are not efficient line emitters. So, the shocks signatures are easily observed where the gas is not photo-ionized, producing the observed morphology. Alternatively, the higher gas excitation perpendicularly the ionization cone can be interpreted as an being produced by shocks due to equatorial AGN winds, as suggested by theoretical recent outflowing torus models \citep[e.g.][]{elitzur12}.

 For H$_2$, we find that the emission lines are narrower  than the [Fe\,{\sc ii}] lines (e.g., Figs.~\ref{linefits} and \ref{maps_subnarrow}), indicating that they do not originate in the same regions, in spite of the strong correlation between the H$_2$ and [Fe\,{\sc ii}] line luminosities observed in samples of ultra-luminous infrared galaxies \citep{hill14} and nearby AGN \citep{rogerio06} using single aperture spectra. One possibility to explain both the correlation between the line luminosities in previous works and the differences between the  H$_2$ and [Fe\,{\sc ii}] line line widths is that H$_2$ is dissociated by shocks (or AGN radiation) in the region where the [Fe\,{\sc ii}] emission is produced, but it is excited in a region very close to the [Fe\,{\sc ii}] emission zone. Assuming that the [Fe\,{\sc ii}] emission arises from a lower density gas, close to the cloud edges, and the H$_2$ from a region with higher density and more distant from the cloud edges, one would expect lower line widths for the H$_2$ and a correlation between the H$_2$ and [Fe\,{\sc ii}] fluxes, if the lines are excited in the same way.

 Another striking kinematic component revealed by our analysis is the broad component ($W_{\rm 80}\gtrsim1000\,$km\,s$^{-1}$)  observed in all emission lines and fitted by two Gaussian functions, one blueshifetd by $\sim150-200$\,km\,s$^{-1}$ and the other redshifted by similar velocities. 
Unlike what is observed for the narrow component, the flux distributions for the broad component do not show any preferred orientation: all lines present a round, centrally peaked flux distributions (see Fig.\ref{maps_subnarrow}). 

The $W_{\rm 80}$ maps for the broad component  (Fig.\ref{maps_subnarrow}) can be used to investigate its origin. Values larger than 1\,000 km\,s$^{-1}$ are observed at all locations, with the highest values of up to 2\,000 km\,s$^{-1}$ seen for the [Fe\,{\sc ii}]$\lambda$1.2570\,$\mu$m to the northeast of the nucleus. 
For a pure Gaussian line profile, $W_{\rm 80}= 2.563\times\sigma$ \citep{zakamska14}. For NGC\,1275, we derive a stellar velocity dispersion of $\sigma_\star =265\pm26$\,km\,s$^{-1}$ (see Fig.\ref{stelfit}), which corresponds to $W_{\rm 80}=680\pm66$\,km\,s$^{-1}$. As the gas shows $W_{\rm 80}$ values much larger than the stellar $W_{\rm 80}$, the kinematics of the broad component cannot be explained by gas motions due to the gravitational potential of the galaxy. High values of $W_{\rm 80}$ observed in ionized gas emission lines are  commonly interpreted as being due to AGN driven winds \citep[e.g.][]{wylezalek20}, which is the most likely scenario for NGC\,1275 as it hosts a strong AGN -- $L_{\rm bol}\approx10^{45}\,{\rm erg\,s^{-1}}$ \citep{woo02}. These outflows may be responsible for providing the shocks necessary to produce the observed line emission of H$_2$ and [Fe\,{\sc ii}], as mentioned above.  

 \subsection{Geometry of the outflows}

The broad-line component presents a round, centrally peaked flux distribution and the velocities of the blueshifted and redshifted components are almost constant over the NIFS field of view (see middle panels of Figs.~\ref{FeIImg}, \ref{Pabmg} and \ref{H2mg}). 
The round emission-line flux distributions suggest that the outflows present a spherical geometry. Our $W_{\rm 80}$ maps show the highest values  surrounding the nucleus at 0\farcs5--1\farcs0 and the $W_{\rm 80}$ maps do not show a smooth variation as seen in the emission-line flux distributions. However, in the inner $0\farcs$5 radius there is a contribution of the compact rotating disk, which may decrease the $W_{\rm 80}$ in these region and the outflow may interact with the gas of the disk producing knots of enhanced $W_{\rm 80}$ values.
From the multi-Gaussian fit, we find that in the inner  $0\farcs$5 radius the redshifted component to the northeast and the blueshifted component to the southwest may be due to the emission of the compact disk. These components seem to be outshining the outflow component from these locations. The most prominent structure in the  $W_{\rm 80}$ map for the [Fe\,{\sc ii}] (Fig.~\ref{maps_subnarrow}) is the region of larger values to the northeast of the nucleus. This structure is approximately along the orientation of the disk and a possible interpretation is that it is produced by the interaction of the outflows with the gas of the disk. Thus, the spherical geometry seems to be consistent with both the  emission-line flux distribution and kinematics observed in the inner region of NGC\,1275. In addition, this geometry is consistent with theoretical studies and numerical simulations of the AGN driven winds that produce sub-relativistic, wide-angle winds \citep[e.g.,][]{silk98,ramos-almeida17,giustini19}.

In addition, the H$_2$ presents overall lower  $W_{\rm 80}$ values than the [Fe\,{\sc ii}]. A possible explanation for this behaviour is that the [Fe\,{\sc ii}] emission is produced in a region closer to the shock front, while the H$_2$ emission would be produced in a shell of denser gas outwards. This is consistent with shock models used to describe the H$_2$ molecule formation from H$_2$ luminous  galaxies, in which the post-shock medium is observed to be multiphase, with H$_2$ gas coexisting with a hot X-ray emitting plasma,  \citep[e.g.,][]{guillard09} as observed in NGC\,1275 \citep[e.g][]{sanders16}.

\subsection{Power of the outflows and implications}

From the observed gas kinematics we can derive the mass-outflow rate and  kinetic power of the outflows. The mass-outflow rate can be determined by  
\begin{equation}\label{eqmout}
\dot{M}^{\rm out}_{\rm ion}=1.4\,m_p\,N_e\,v_{\rm out}\,f\,A ,
\end{equation}
where $m_p$ is the proton mass, $N_e$ is the electron density, $v_{\rm out}$ is the velocity of the outflow, $f$ is the filling factor, $A$ is the area of the outflow cross section and the factor 1.4 is to account for elements heavier than H. The filling factor can be estimated by
\begin{equation}
L_{\rm Pa\beta}= 4\pi j_{\rm Pa\beta}\,V\,f, 
\end{equation}
where $L_{\rm Pa\beta}$ is the Pa$\beta$ luminosity emitted by a volume $V$ and $j_{\rm Pa\beta}$ is the Pa$\beta$ emission coefficient. For the low-density limit and a temperature of 10\,000\,K,  $4\pi j_{\rm Pa\beta}/N_e^2 = 2.046\times10^{-26}$  erg\, cm$^{-3}$\,s$^{-1}$ \citep{oster06}. Replacing $f$ in the equation for $\dot{M}^{\rm out}_{\rm ion}$, we obtain 

\begin{equation}
\label{dotm}
\dot{M}^{\rm out}_{\rm ion}=\frac{1.4\,m_p\,N_e}{4\pi j_{\rm Pa\beta}}\,\frac{A}{V}\,L_{\rm Pa\beta}\,v_{\rm out}.
\end{equation}

To estimate $\dot{M}^{\rm out}_{\rm ion}$ we assume that the outflows are spherical with a radius of 0\farcs9 ($r_{\rm out}=270$\,pc), defined as the radius that enclose 90\,\% of the flux of the Pa$\beta$ emission line. A similar definition was used by \citet{fischer18} and it is worth mention that it is similar to the value (200\,pc) derived by \citet{baron19} for the peak of the distribution of the location of outflows in 1700 type 2 AGN based on modeling of their spectral energy distribution.  We adopt $N_e=500$\,cm$^{-3}$, a typical value of the electron density measured for AGN based on the [S\,{\sc ii}] emission lines \citep[e.g., ][]{dors14} and  $L_{\rm Pa\beta} = 5.4\times10^{39}$ erg\,s$^{-1}$, by integrating the fluxes of the broad Pa$\beta$ components. Following previous spatially resolved studies of outflows \citep{rogemarMrk79,diniz19,shimizu19,munoz-vergara19,soto-pinto19,fischer19,gnilka20}, we use the measured centroid velocity of the broad components as the velocity of the outflow: $v_{\rm out}=200$\,km\,s$^{-1}$ (see middle panels of Figs.~\ref{FeIImg}, \ref{Pabmg} and \ref{H2mg}). This velocity traces the bulk of the outflowing gas. Under these assumptions we obtain $\dot{M}^{\rm out}_{\rm ion}\approx1.6$\,M$_\odot$\,yr$^{-1}$ in ionized gas. 
 The bolometric luminosity of the AGN in NGC\,1275 is $\sim1.1\times10^{45}$\,erg\,s$^{-1}$ \citep{woo02} and with the derived $\dot{M}^{\rm out}_{\rm ion}$, NGC\,1275 falls very close to the wind scaling relations for ionized gas in AGN \citep{fiore17,shimizu19}. 

As we observe similar ``bulk" velocities for the molecular and ionized gas outflows (although the ionized gas traces more turbulent clouds, as indicated by the higher $W_{\rm 80}$ values), we can estimate the outflow in hot molecular gas by multiplying $\dot{M}^{\rm out}_{\rm ion}$ by the ratio of the masses of hot molecular and ionized gas. The total mass of ionized gas in the outflow can be estimated by
\begin{equation} 
M_{\rm ion} \approx 6.0\times 10^{8}\left(\frac{L_{\rm Pa\beta}}{\rm 10^{40} erg\,s^{-1}}\right) \left(\frac{N_e}{\rm cm^{-3}}\right)^{-1} M_\odot,
\end{equation}
where $L_{\rm Pa\beta}$ is the luminosity of the region that is in outflow
As in Eq.\,\eqref{dotm}, we assume the low-density limit and a temperature of 10\,000\,K and include a factor 1.4 to account for heavier elements \citep{scoville82,oster06,sb09}. Using the parameters above, we estimate  $M_{\rm ion}\approx6.5\times10^5\,{\rm M_\odot}$.  
The mass of hot molecular gas is given by
\begin{equation}\label{H2_mass} 
M_{\rm H_2} \approx 4.24\times 10^{3}\left(\frac{L_{\rm H_2}}{\rm 10^{40} erg\,s^{-1}}\right) {\rm M_\odot},
\end{equation}
where $L_{\rm H_2}$ is the luminosity of the H$_2\lambda2.1218\,\mu$m emission line and this equation is derived under the assumption that the H$_2$ emitting gas is in thermal equilibrium \citep{scoville82,rogemar_n4051,schonell19}. Integrating the flux of the broad H$_2$ emission over the NIFS field of view, we estimate $L_{\rm H_2} = 2.6\times10^{40}$ erg\,s$^{-1}$ and thus $M_{\rm H_2}\approx1.1\times10^4\,{\rm M_\odot}$. 

The mass outflow rate in hot molecular gas is $\dot{M}^{\rm out}_{\rm H_2}\approx\dot{M}^{\rm out}_{\rm ion}\times\frac{M_{\rm H_2}}{M_{\rm ion}}=2.7\times10^{-2}\,{\rm M_\odot}$. In nearby AGN ($z<0.02$), the hot molecular gas emission is usually dominated by gas rotating in the galaxy disk \citep{davies14,rogemar_llp_sample,sb19}, while molecular outflows are still scarce \citep{rogemar_eso,rogemar_N5929_letter,rogemar_N5929_paper,diniz17,may17,shimizu19} and the  $\dot{M}^{\rm out}_{\rm H_2}$ derived for NGC\,1275 is in the range of  values in these previous studies.  

However, the hot molecular gas represents only a small fraction of the total amount of molecular gas -- the amount of cold molecular gas is usually $10^5-10^7$ times larger than that of hot molecular gas \citep{dale05,ms06,mazzalay13}. For the disk component, we estimate a mass of hot molecular gas of 3500\,M$_\odot$ using the integrated flux of H$_2$\,$\lambda$2.1218\,$\mu$m for the narrow component (Fig.\,\ref{maps_subnarrow}) and Eq.\,\ref{H2_mass}, while mass of cold molecular gas in the disk is $4\times10^8$\,M$_\odot$ \citep{nagai19}. Thus, the ratio between the masses of cold and hot molecular gas in the compact disk of NGC\,1275 is in agreement with the measurements for other galaxies. On kpc scales, NGC\,1275 is known to present large amount (10$^{10}$\,M$_\odot$) of molecular gas \citep{salome06,lim08}.  If the outflow component presents a similar relation between the amount of hot and cold molecular gas than that for the disk component, the total outflow rate in molecular gas in NGC\,1275 may be much larger than the value estimated above and compatible with those derived for powerful AGN and ultra luminous infra-red galaxies of up to 10$^3$\,M$_\odot$\,yr$^{-1}$ \citep[e.g.,][]{veilleux13,lutz20}. Such outflows are massive and powerful enough to provide significant AGN feedback as requested by numerical simulations \citep{dimatteo05,harrison17} to describe the evolution of galaxies. 
Indeed, ALMA observations of NGC\,1275 reveal fast molecular outflows as absorption features in the dense gas tracers HCN(3--2) and HCO$^+$(3--2), while the CO(2--1) emission is dominated by the compact molecular disk \citep{nagai19}, which could be outshining the outflowing material. Alternatively, it is possible that the outflowing molecular gas is shock-heated to temperatures much higher than the typical ISM values, in which case the hot-to-cold ratio for the outflow could be atypically large and therefore the cold molecular gas outflow rate could be small.

From the mass-outflow rate and the observed kinematics, we can derive the kinetic power of the outflows by $\dot{E}_{\rm out}=\frac{1}{2}\dot{M}^{\rm out}_{\rm ion} (v_{\rm out}^2 + 3\,\sigma_{\rm out}^2)$, where $\sigma_{\rm out}$ is the velocity dispersion of the outflow, which can be estimated by $\sigma_{\rm out}=W_{\rm 80}/2.563$. Adopting $W_{\rm 80}=1500\,$km\,s$^{-1}$, a median value $W_{\rm 80}$ for the [Fe\,{\sc ii}]$\lambda1.2570\,\mu$m (Fig.\,\ref{maps_subnarrow}), we obtain  $\dot{E}_{\rm out}=5.4\times10^{41}$ erg\,s$^{-1}$. This value is only 0.05 per cent of the AGN bolometric luminosity and thus, the ionized outflows are not powerful enough to affect the evolution of the galaxy, as outflows with kinetic powers of at least 0.5\,\% of the AGN luminosity are effective in suppressing star formation in the host galaxy \citep{hopkins10}. On the other hand, if we extrapolate our result for the hot molecular outflow to the cold molecular gas, the kinetic power of the cold molecular outflows would be a few per cent of the AGN bolometric luminosity and thus being powerful enough to affect the evolution of the galaxy. In addition, numerical simulations show that the kinetic energy corresponds to less than 20 per cent of the total energy of the outflow \citep{richings18b}, and thus the total power of the outflow in NGC\,1275 is at least 5 times larger than the values derived above.

NGC\,1275 is a massive, cD galaxy, with a stellar mass of $\sim2.43\times10^{11}$\,M$_\odot$  and an effective radius of 6.41 kpc \citep{mathews06}. Assuming that the stellar mass is 15 per cent of the total mass of the galaxy, we estimate the escape velocity to be $\sim 1500$ km s$^{-1}$. Thus, the velocities of the outflows in NGC\,1275 are smaller than its escape velocity, implying in a  ``maintenance mode" feedback, where the outflows re-distribute the gas within the galaxy, but it still remains available for further star formation.  In addition, our data reveal that shocks due to the AGN winds in NGC\,1275 play an important role in the excitation of the observed H$_2$ and [Fe\,{\sc ii}] emission.

\section{Conclusions}

We use Gemini NIFS adaptive optics observations to map the near-IR emission-line flux distribution and kinematics of the inner $\sim$500 pc radius of NCG\,1275 at a spatial resolution of $\sim$70 pc. Our main conclusions are:

\begin{itemize}
    \item From the fitting of the K-band CO absorption features, we derive a stellar velocity dispersion of $\sigma_\star =265\pm26$\,km\,s$^{-1}$ (within a ring with inner radius of 0\farcs75 and outer radius of 1\farcs25), which represents the first measurement of $\sigma_\star$ based on near-IR spectra. Using the $M_{\rm SMBH}-\sigma_\star$ relation we estimate a mass for the SMBH of  $M_{\rm SMBH}=1.1^{+0.9}_{-0.5}\times10^9$\,M$_\odot$, in good agreement with previous dynamical measurements.
    
    \item The emission-line profiles of the molecular and ionized gas present multi-kinematical components: one narrow $(\sigma<150$ km\,s$^{-1}$), attributed to gas orbiting in the galaxy disk and two broad $(\sigma>150$ km\,s$^{-1}$), one blueshifted and another redshifted, attributed to an spherical outflow.
    
    \item The flux distributions of the narrow-line component ($\sigma<150$\,km\,s$^{-1}$) of the ionized and molecular gas are distinct. While the ionized gas emission is  observed surrounding the nucleus and extending along the AGN ionization axis (PA=140/320$^\circ$), the molecular emission is observed perpendicularly to it, following the orientation of the compact disk previously observed in the inner 100 pc.  
    
    \item The narrow-line component presents velocities close to the systemic velocity of the galaxy in all locations. Combined with the corresponding flux distributions, lead to the interpretation that this component is tracing the emission of an almost face on disk, perpendicular to the AGN axis. The AGN radiation seems to be responsible for the dissociation of the H$_2$ molecule in locations close to the ionization axis.
    
    \item The outflow is seen as broad line components with velocities of up to 2\,000 km\,s$^{-1}$. The emission-line flux distribution of outflowing gas is round and centrally peaked and is well reproduced by two Gaussian-functions, one blueshifted by 150--200 km\,s$^{-1}$ and the other redshifted by similar velocities, relative to the systemic velocity of the galaxy. 
    
    \item From the emission line-ratios we find that shocks originated in AGN winds plays are the dominant production mechanism of the observed line emission in the inner $\sim$500 pc, while a small contribution of AGN radiation is seen along the ionization axis.
    
    \item We derive a mass outflow-rate in ionized gas of 1.6\,M$_\odot$\,yr$^{-1}$ and a kinetic power of $2\times10^{40}$ erg\,s$^{-1}$. In hot molecular gas, we find outflows at $2.7\times10^{-2}\,{\rm M_\odot}$\,yr$^{-1}$, which could represent only a small fraction of the total molecular outflows, possibly dominated by colder gas phases.
    
    \item The velocities and kinetic powers of the hot molecular ($\dot{E}{\rm out}\approx8\times10^{-6} L_{\rm AGN}$) and ionized ($\dot{E}{\rm out}=5\times10^{-4} L_{\rm AGN}$) outflows are not high enough to provide significant AGN feedback. The outflows are only able to redistribute the gas in the galaxy. 
  
\end{itemize}

\section*{Acknowledgements}
We thank an anonymous referee for useful suggestions which
helped to improve the paper.  RAR, TSB and RR thank CNPq, FAPERGS and CAPES for finacial supporting. This study was financed in part by Conselho Nacional de Desenvolvimento Cient\'ifico e Tecnol\'ogico (202582/2018-3, 304927/2017-1 and 400352/2016-8) and Funda\c c\~ao de Amparo \`a pesquisa do Estado do Rio Grande do Sul (17/2551-0001144-9 and 16/2551-0000251-7). Based on observations obtained at the Gemini Observatory, which is operated by the Association of Universities for Research in Astronomy, Inc., under a cooperative agreement with the NSF on behalf of the Gemini partnership: the National Science Foundation (United States), National Research Council (Canada), CONICYT (Chile), Ministerio de Ciencia, Tecnolog\'{i}a e Innovaci\'{o}n Productiva (Argentina), Minist\'{e}rio da Ci\^{e}ncia, Tecnologia e Inova\c{c}\~{a}o (Brazil), and Korea Astronomy and Space Science Institute (Republic of Korea). 
This research has made use of NASA's Astrophysics Data System Bibliographic Services. This research has made use of the NASA/IPAC Extragalactic Database (NED), which is operated by the Jet Propulsion Laboratory, California Institute of Technology, under contract with the National Aeronautics and Space Administration. The Pan-STARRS1 Surveys (PS1) and the PS1 public science archive have been made possible through contributions by the Institute for Astronomy, the University of Hawaii, the Pan-STARRS Project Office, the Max-Planck Society and its participating institutes, the Max Planck Institute for Astronomy, Heidelberg and the Max Planck Institute for Extraterrestrial Physics, Garching, The Johns Hopkins University, Durham University, the University of Edinburgh, the Queen's University Belfast, the Harvard-Smithsonian Center for Astrophysics, the Las Cumbres Observatory Global Telescope Network Incorporated, the National Central University of Taiwan, the Space Telescope Science Institute, the National Aeronautics and Space Administration under Grant No. NNX08AR22G issued through the Planetary Science Division of the NASA Science Mission Directorate, the National Science Foundation Grant No. AST-1238877, the University of Maryland, Eotvos Lorand University (ELTE), the Los Alamos National Laboratory, and the Gordon and Betty Moore Foundation.




\bibliographystyle{mnras}
\bibliography{N1275_r1} 



\appendix\section{Multi-Gaussian Fits of the Emission-Line Profiles}\label{appendix}

\begin{figure*}
    \centering
    \includegraphics[width=0.95\textwidth]{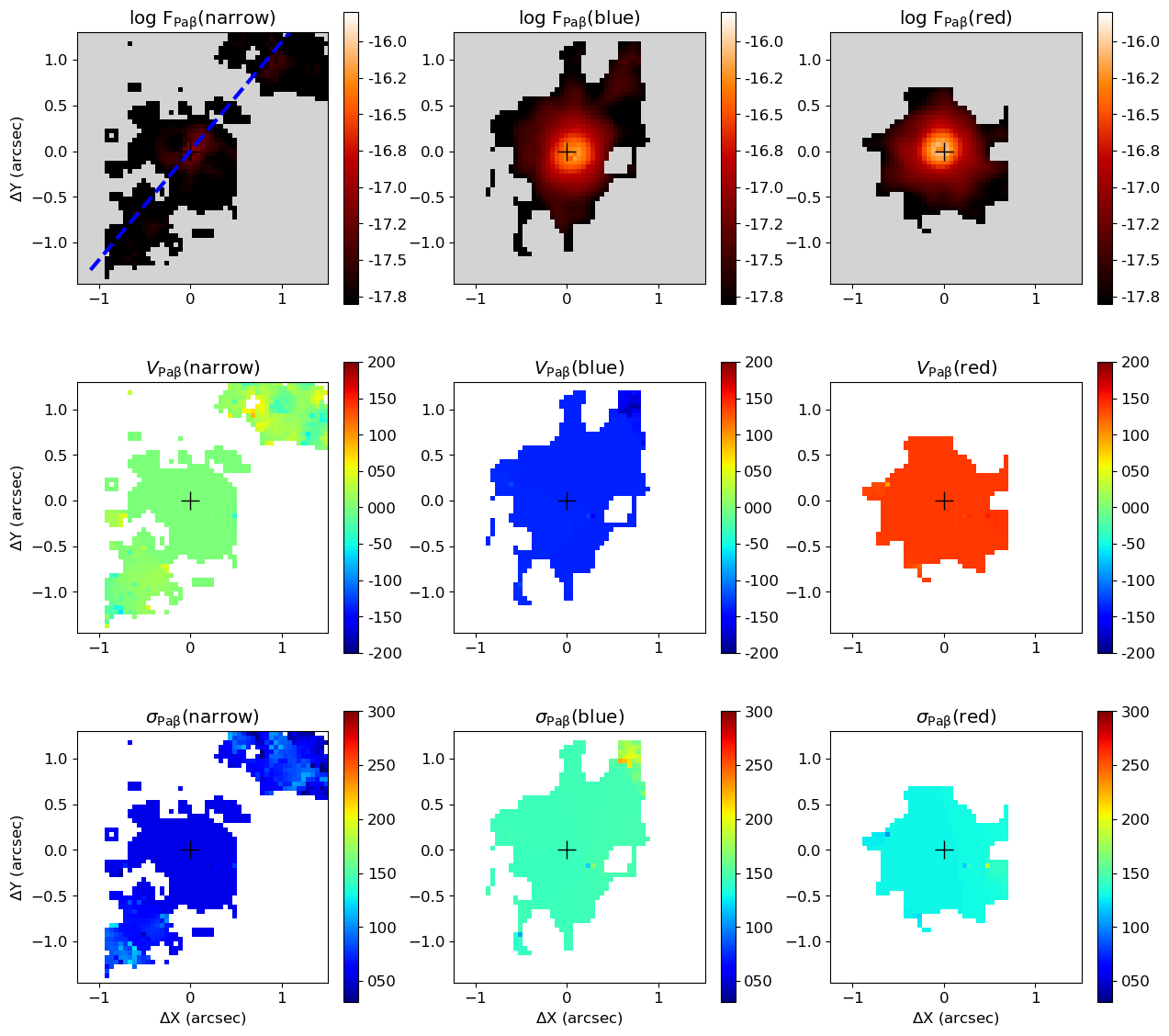}
    \caption{Flux, velocity and velocity dispersion maps from multi-Gaussian components fits of the Pa$\beta$ emission-line profile. The dashed line shows the orientation of the collimated Pa$\beta$ emission.}
    \label{Pabmg}
\end{figure*}

\begin{figure*}
    \centering
    \includegraphics[width=0.95\textwidth]{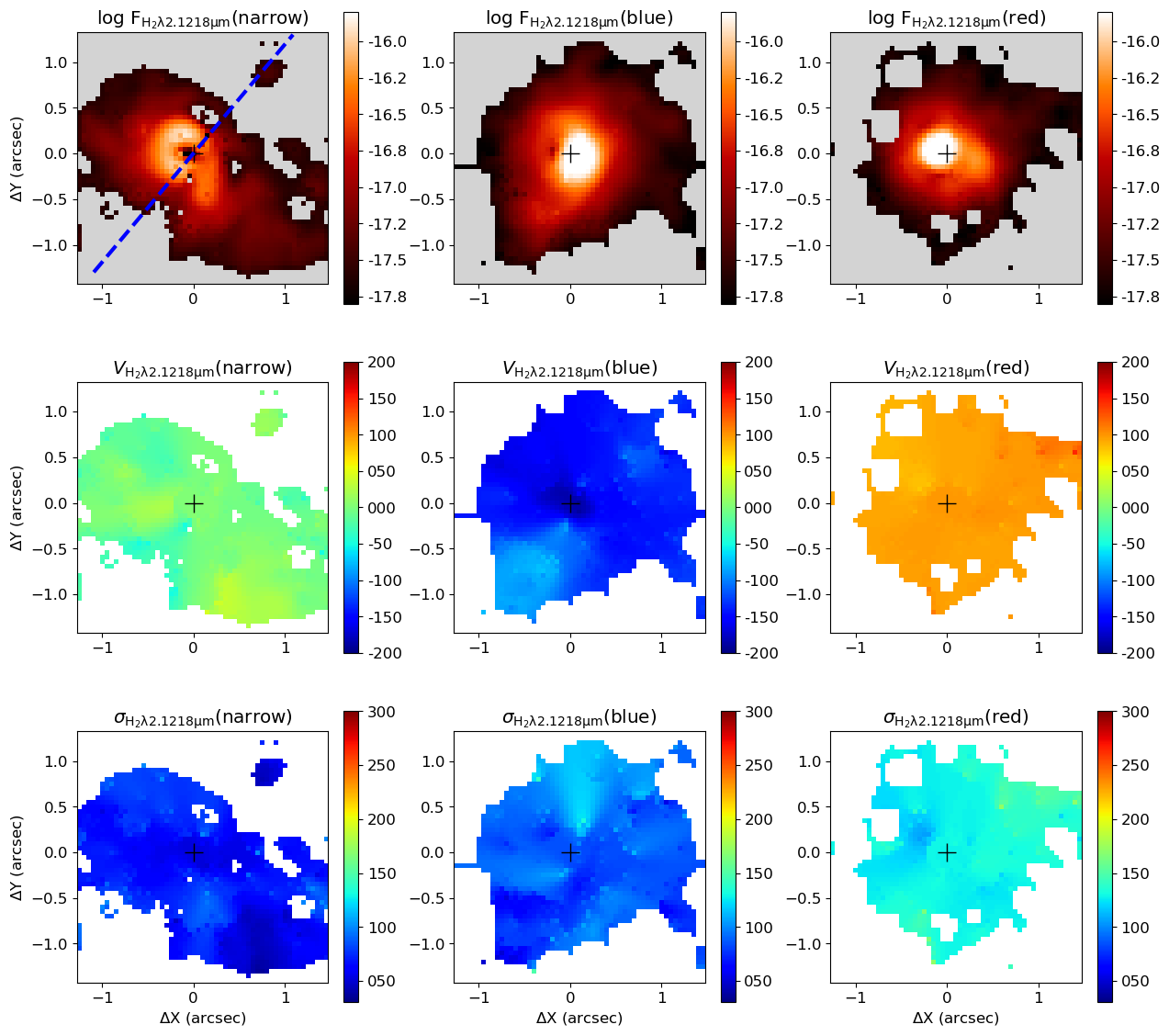}
    \caption{Same as Fig.~\ref{Pabmg} for the  H$_2\lambda2.1218\,\mu$m emission line. }
    \label{H2mg}
\end{figure*}


\bsp	
\label{lastpage}

\end{document}